\documentclass[aip,amsmath,amssymb,reprint]{revtex4-1}

\input{myPreamble.tex}

\makeatother

\begin{document}

\preprint{AIP/123-QED}

\title{Effect of filter kernel on scale-energetics of near-wall turbulent structures}

\author{Daniel~Feldmann}\email{daniel.feldmann@uni-bremen.de}
\affiliation{Universität Bremen, Institute for Integrated Product Development (BIK),\\Badgasteiner Straße 1, 28359 Bremen, Germany.}
\affiliation{Universität Bremen, Center of Applied Space Technology and Microgravity (ZARM),\\Am Fallturm 2, 28359 Bremen, Germany.}
\author{Mohammad~Umair}
\affiliation{Universität Bremen, Center of Applied Space Technology and Microgravity (ZARM),\\Am Fallturm 2, 28359 Bremen, Germany.}
\affiliation{Laboratoire des Écoulements Géophysiques et Industriels (LEGI), UMR 5519, CNRS,\\BP 53, 38041 Grenoble CEDEX 09, France}
\author{Marc~Avila}
\affiliation{Universität Bremen, Center of Applied Space Technology and Microgravity (ZARM),\\Am Fallturm 2, 28359 Bremen, Germany.}
\affiliation{Universität Bremen, MAPEX Center for Materials and Processes,\\Am Biologischen Garten 2, 28359 Bremen, Germany.}
\author{Alexandra~von~Kameke}
\affiliation{Hamburg University of Applied Sciences, Department of Mechanical Engineering and Production Management,\\Berliner Tor 21, 20099 Hamburg, Germany.}
\date{\today}

\begin{abstract}
Inter-scale energy fluxes, $\Pi^\lambda$, are widely used as a diagnostic tool to analyse energy transfer across length scales, $\lambda$, in turbulence data. Here, we investigate how the choice of filter kernel (sharp spectral, Gaussian, box) affects the computed energy fluxes at constant filter width. We apply spatial filtering to a turbulent pipe flow simulation dataset and assess the effect on the local structure of $\Pi$. While the mean energy flux profile at each wall-normal distance is qualitatively robust across kernels, we observe significant differences in the intensity and spatial distribution of localised $\Pi$ events. Correlations between typical flow structures in the buffer layer (streaks, vortices, and Q-events) and regions of forward/backward transfer in the instantaneous $\Pi$ field differ markedly between kernel types. Cross-correlations appear strongly upstream--downstream symmetric when using the sharp spectral kernel, but asymmetric for the Gaussian and box kernels. For the Gaussian and box kernels $\Pi$ events tend to localise along the inclined meander of streaks, while they are centred on top of the streaks for the sharp spectral kernel. Moreover, using the sharp spectral kernel, we observe a coincidence of backward scatter and fluid transport away from the wall ($Q_1$), which does not appear with the Gaussian and box kernels. All kernels, however, predict backward scatter directly downstream of $Q_1$ events. The results suggest that interpretations of inter-scale energy flux based on sharp spectral scale separation should be treated with caution, since such kernels act non-local in physical space, whereas $\Pi$ events are inherently localised. Our python post-processing tool \eflux for scale separation and energy flux analysis in pipe flows is freely available and readily adaptable to other flow configurations and filter widths.
\end{abstract}


\maketitle

\section{Introduction}
\label{sec:introduction}

Spatial or temporal coarse-graining is an attractive approach to separate scales and has become a central tool for the analysis of turbulent flows \cite{Leonard1975, Germano1992, Liu1994, Tao2002, Bai2013, Drivas2017, Bian2019, Araki2024, Zhou2024, Park2025}. It is often achieved by applying an ideal low-pass filter to velocity datasets \cite{Piomelli1996, Miller2014, Buzzicotti2018a, Bauer2019, Dogan2019, Kawata2021, Kannadasan2024}, albeit the suitability of such a sharp spectral cut-off has occasionally been questioned \cite{Eyink1994, Eyink1995, Mishra2014}.

Separating larger from smaller scales has helped to uncover many details about the energy cascade and the formation of structures in turbulent flows. Especially the study of turbulent superstructures in wall-bounded systems \cite{Smits2011} has renewed interest in scale separation in order to differentiate the share of the larger scales on the turbulent kinetic energy budget and the Reynolds stresses \cite{Mizuno2016, Ahn2017, Bauer2019, Lee2019}. Increasingly often, the filtered flow field is used to compute the inter-scale flux, $\Pi^\lambda$, of turbulent kinetic energy through a particular filter length scale, $\lambda$, based on a framework formulated by \textcite{Eyink1995} or to compute other, very similar inter-scale energy transfer markers \cite{Cardesa2019}.

In homogeneous isotropic turbulence (HIT), the interpretation of $\lambda$ and $\Pi$ is straightforward and has been studied in detail by \textcite{Cardesa2017}. The locality assumption of the energy flux in wavenumber space -- as postulated by \textcite{Richardson1922} and \textcite{Kolmogorov1941} -- has been proven theoretically \cite{Eyink2009, Aluie2009, Eyink2005, Eyink1995} and also demonstrated empirically for homogeneous shear flow and isotropic turbulence \cite{Cardesa2015}.

\begin{figure*}
\includegraphics[width=1.0\textwidth, trim=1.5mm 0mm 0mm 0mm, clip=true]{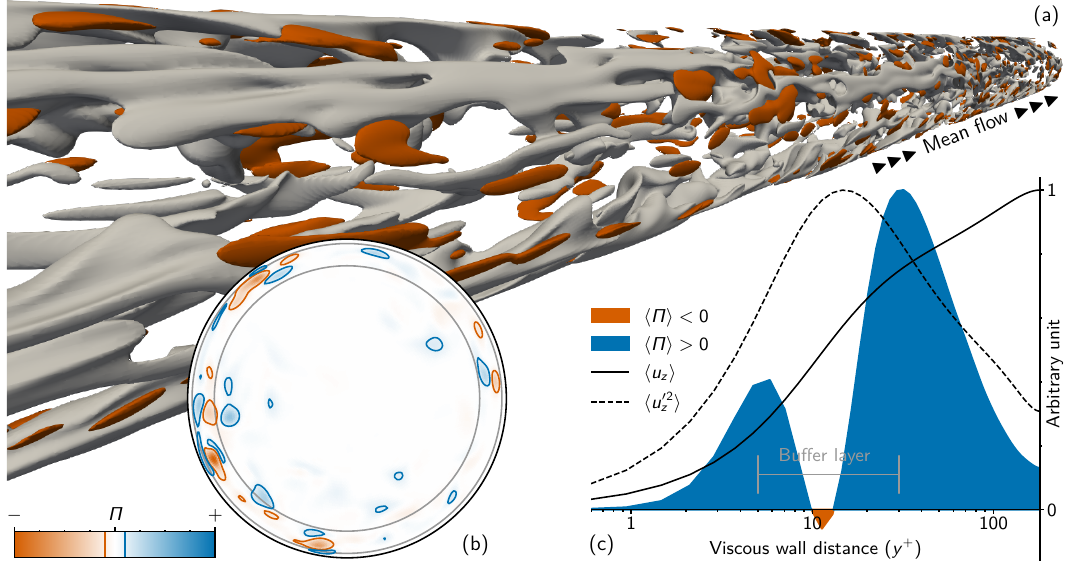}
\caption{Overview of our wall-turbulence data set (\ie pipe flow DNS at Reynolds number $\ReTau=\num{180}$ in a computational domain of length $L=\num{42}R$) and the inter-scale energy flux, $\Pi$, computed from that dataset based on a Gaussian filter kernel. (a): Large connected regions of negative streamwise velocity fluctuations (\textcolor{Grey}{grey iso-contours} for $u^{\prime}_z = \num{-2.5}\utau$) represent typical low-momentum streaks in the near-wall region of the pipe domain. The streaks appear accompanied by shorter regions of instantaneous backward flux (\textcolor{Vermillion}{red iso-contours} for $\Pi=\sfrac{-1}{10}\max\lvert\Pi\rvert$). (b): Forward (\positive) and backward (\negative) scatter events in a cross-sectional ($r$--$\theta$) plane, that contains the most intense backscatter event occurring in this instantaneous flow field realisation. Significant events appear clustered in the buffer layer (\textcolor{Grey}{annular region between grey lines}). (c): Mean flux as function of the distance from the pipe wall. A distinct region of net small-to-large-scale energy flux, \ie $\langle\Pi\rangle<\num{0}$, indicates a localised inverse energy cascade in the buffer layer ($\num{5}\le y^{+}\le\num{30}$). Following the convention in other canonical flows, we use $y^+$ to indicate the distance from the wall in viscous units. The mean streamwise velocity, $\langle u_z\rangle$, and streamwise Reynolds stress, $\langle u^{\prime 2}_z\rangle$, serve as reference for regions of maximal mean shear and peak turbulence intensity.}
\label{fig:streaksAndFlux}
\end{figure*}

Wall-bounded turbulence, however, is inhomogeneous and anisotropic by nature. As an example, figure~\ref{fig:streaksAndFlux}a shows elongated low-momentum streaks populating the buffer layer in a turbulent pipe flow extracted from our direct numerical simulation (DNS) data. Streamwise streaks are probably the most prominent and best investigated structural features of near-wall turbulence, first described by \textcite{Kline1967}. Together with quasi-streamwise vortices, they play a major role in the self-sustaining wall cycle \cite{Jimenez1999} and they are often implicitly associated with sweep ($Q_4$) and ejection ($Q_2$) events \cite{Wallace2016}. In wall-bounded systems, the multi-dimensional energy fluxes are more elusive and exhibit much richer physics compared to HIT. This was shown in a series of studies of the scale energetics in a channel flow using an alternative approach for scale separation based on the generalised Kolmogorov equation (GKE) \cite[\eg][]{Marati2004, Saikrishnan2012, Cimarelli2013, Cimarelli2016}. In contrast to the statistical GKE approach, $\Pi$ is a dynamic quantity depending on time and all three spatial dimensions. This allows for a pointwise comparison of the local structure and temporal dynamics of patterns appearing in the $\Pi$ field to individual realisations of localised structures appearing in the turbulent velocity field (fig.~\ref{fig:streaksAndFlux}a).

The study of $\Pi$ in wall-bounded systems stems from considerations on model and discretisation errors in large eddy simulations (LES) \cite{Germano1992}. A set of important early work aimed at better understanding and modelling the sub-grid scale (SGS) processes in LES using a priori assessment of the relevant SGS quantities based on DNS data \cite{Piomelli1991, Piomelli1996, Domaradzki1993, Domaradzki1994, Haertel1994, Kerr1996}. These studies generated much physical insight into the scale-energetics of the wall cycle. Next to \citet{Meneveau1991a, Meneveau1991b}, they were the first to report the existence of so-called backscatter events in the buffer layer of wall-bounded turbulent flows (figs.~\ref{fig:streaksAndFlux}a and b) and they showed that the inter-scale transfer of energy is highly promoted by strong shear layers (fig.~\ref{fig:streaksAndFlux}c). In fact, \textcite{Haertel1994} report a negative net energy flux across a range of scales indicating a localised inverse energy cascade in the buffer layer, where smaller scales pass their energy to larger ones on the average (fig.~\ref{fig:streaksAndFlux}c). Additionally, the studies of \textcite{Haertel1994} and \textcite{Piomelli1996} consider the coincidence of forward ($\Pi>\num{0}$) and backward ($\Pi<\num{0}$) scatter events with conditional mean flow structures and other quantities derived from the turbulent velocity field, such as regions of strong strain.

\textcite{Haertel1994} used a variable interval space average analysis and a threshold to connect strong wall-normal velocity gradients with strong local energy flux events. However, with their analysis method they were not able gain insight into the spanwise topography of the $\Pi$ field. Expanding on these findings, \textcite{Piomelli1996} performed flow field averaging conditioned to strong forward or backward scatter events. Thus, they were able to connect the presence of forward and backward scatter to the upwash and downwash side of streamwise vortices visible in the conditional mean flow. From that they concluded that close to the wall (at $y^+=\num{14}$) forward scatter is predominantly related to fluid transport away from the wall (via $Q_2$ events), while backward scatter occurs mostly along with fluid transport towards the wall (via $Q_4$ events); by now a common view of the near-wall cycle. Note, that $y^+$ indicates wall-normal distances measured in viscous units, as we define later. In both studies, the computation of the inter-scale energy flux relies on a two-dimensional (2D) sharp spectral cut-off to separate length scales in turbulent velocity fields generated by DNS.

In this paper, we show that cross-correlations of typical turbulent near-wall features (streaks, vortices, $Q$ events) and the inter-scale energy flux highly depend on the type of filter kernel employed for scale separation. In section~\ref{sec:methodology} we briefly introduce our pipe flow DNS data, the concept of inter-scale energy transfer, and the applied filtering techniques, while details are left for the appendix. Our results based on the commonly used sharp spectral kernel are discussed in section~\ref{sec:resultsFourier} and compared to the existing literature in terms of instantaneous snapshots and one- and two-point statistics. The differences in the inter-scale energy flux and its cross-correlations with streaks, vortices and $Q$ events for different types of kernels are detailed in section~\ref{sec:filterEffect}. In section~\ref{sec:discussion}, we discuss the implications of our findings for the interpretation of the scale energetics of near-wall turbulence structures as well as the limitations of our study.
 
\section{Methodology}
\label{sec:methodology}

\subsection{DNS dataset}
\label{sec:dnsDataBase}

In order to study the effect of the different filter kernels on the local structure of the scale-energetics in wall-bounded turbulence, we generated a well-resolved velocity data set of a turbulent pipe flow at a Reynolds number of $\ReTau=\sfrac{R\,\utau}{\nu}=\num{180}$; the same value as previously used by \textcite{Haertel1994} and \textcite{Piomelli1996}. Here, $R=\sfrac{D}{2}$ is the pipe radius, \utau is the friction velocity, and $\nu$ is the kinematic viscosity of the fluid.

The data that supports this study is available at \pangaea and is computed from \num{351} full velocity snapshots taken from a DNS we performed with our publicly available simulation code \nsp \cite{Lopez2020}. In \nsp, the incompressible Navier--Stokes equations are formulated in cylindrical coordinates ($r,\theta,z$) and integrated forward in time ($t$) based on a pseudo-spectral formulation. The size of the computational domain is $\Omega=\left(R\times\num{2}\pi\times\num{42}R\right)$ in radial ($r$), azimuthal ($\theta$) and streamwise ($z$) direction, respectively, and therefore four times longer than the one used by \textcite{Haertel1994}. The number of radial grid points and Fourier modes used in our DNS is $(N_{r}\times N_{\theta}\times N_{z})=(\num{80}\times\num{128}\times\num{768})$. After dealiasing, this results in a spatial resolution of $\Delta\theta R^{+}=\num{4.4}$ and $\Delta z^{+} = \num{4.9}$, whereas radial grid points are clustered towards the pipe wall such that $\num{0.05}\le\Delta r^{+}\le\num{4.4}$ and \num{24} points lie within the buffer layer ($y^{+}\le\num{30}$). As in other canonical flows, we use $y^+=(R-r)\,\ReTau$ to indicate the distance from the wall and the superscript ${\!}^{+}$ denotes length scales measured in inner units ($\sfrac{\nu}{\utau}$).

Measured in outer units, all generated snapshots cover a time window of $\num{857}\sfrac{D}{\ubulk}$, where \ubulk is the bulk velocity. This is one order of magnitude longer than the persistence of the longest low-level space-time correlations ($\num{20}\sfrac{D}{\ubulk}$) reported by \textcite{Wu2012} in their DNS study focusing on very large scale motions (VLSM) in turbulent pipe flow. Therefore, we expect a sufficiently large uncorrelated statistical sample even for the largest (\ie slowest) scales in the turbulent flow field.

\subsection{Inter-scale flux of turbulent kinetic energy}
\label{sec:eFlux}

Following \textcite{Haertel1994}, we consider the full inhomogeneous and
anisotropic velocity field as composed out of three parts: The statistically
stationary mean flow and the turbulent fluctuations divided into a sub- and a
super-filter part. To compute the flux of turbulent kinetic energy, we therefore
only consider the fluctuating part of the velocity field
\begin{align}
u^{\prime}_{i} = u_{i} - \lla u_{i}\rra_{t,\theta,z}
\quad\text{with}\quad
i\in\{r,\theta,z\}
\text{,}
\label{eq:reynoldsDecomposition}
\end{align}
where angled brackets denote averaging over all available snapshots at different
time instants $t$ and in the two homogeneous spatial directions $\theta$ and
$z$. In the following, we drop these indices for the sake of clarity and also
use the same notation to indicate statistics of other quantities.

For scale separation, the general idea \cite{Leonard1975} is to apply an explicit spatial low-pass filter of the form
\begin{align}
\overline{u^{\prime}_{i}}^{\lambda}(\vec{x},t) =
\int_{\Omega} G^{\lambda}(\vec{x} - \vec{x}^{\ast})\cdot
u^{\prime}_{i}(\vec{x}^{\ast}, t)
\increment\vec{x}^{\ast}
\label{eq:explicitFilter}
\end{align}
to the entire domain $\Omega$, where $G^{\lambda}$ is the filter kernel acting at a nominal filter width $\lambda$. The convolution~\eqref{eq:explicitFilter} yields the super-filter part $\overline{u^{\prime}_{i}}^{\lambda}$, that only contains turbulent fluctuations of length scales larger than approximately $\lambda$; analogously to the resolved scales in the context of LES. The sub-filter part
\begin{align}
\widetilde{u^{\prime}_{z}} & = u^{\prime}_{z} - \overline{u^{\prime}_{z}}
\label{eq:subFilterScales}
\end{align}
represents the removed (residual) fluctuations approximately smaller than the chosen filter width; analogously to the SGS in the context of LES. The second-order
terms $\overline{u^{\prime}_{i}u^{\prime}_{j}}^{\lambda}$ are computed analogously
to eq.~\eqref{eq:explicitFilter}.

Following, \eg, \textcite{Haertel1994}, \textcite{Eyink1995}, \textcite{Kelley2013}, \textcite{Ballouz2018}, the inter-scale flux of turbulent kinetic energy across a length scale $\lambda$ can now be computed as
\begin{align}
\Pi^{\lambda} & = -\tau_{ij} \cdot S_{ij} \\
&= -\left(
\overline{u^{\prime}_{i}u^{\prime}_{j}}^{\lambda} -
\overline{u^{\prime}_{i}}^{\lambda} \overline{u^{\prime}_{j}}^{\lambda}\right)
\cdot\frac{1}{2}\left(\partial_{j}\overline{u^{\prime}_{i}}^{\lambda} +
                      \partial_{i}\overline{u^{\prime}_{j}}^{\lambda}\right)
\text{.}
\label{eq:eFlux}
\end{align}
The first term can be interpreted as a shear stress ($\tau_{ij}$) due to
turbulent fluctuations smaller than $\lambda$, whereas the second term
represents the strain rate ($S_{ij}$) of turbulent flow field patterns larger
than the filter scale. Therefore, $\Pi^{\lambda}$ can be interpreted as a
measure for the work performed by smaller scales to deform (destroy or create)
larger scales. This analogy implies major importance of the spatial alignment of
stress and strain \citep[\eg][]{Ballouz2018}. If $\Pi^{\lambda}>\num{0}$, energy
is transferred across the filter scale $\lambda$ from larger to smaller scales
(\ie forward scatter). If, on the other hand, $\Pi^{\lambda}<\num{0}$, energy is
fed from scales smaller than $\lambda$ to the next larger ones (\ie backward
scatter).

\subsection{Separation of scales}
\label{sec:separationOfScales}

To enable direct comparison with previous studies and to unambiguously isolate the effect of the filter kernel $G$ on $\Pi$ and its statistics, we adopt the same setup as used by \textcite{Haertel1994}. At each wall-normal location $r$, we apply a two-dimensional filter kernel in the homogeneous directions $\theta$ and $z$, using a fixed filter scale of $\left(\lambda^+_\theta \times \lambda^+_z\right) = \left(\num{40} \times \num{75}\right)$. This non-uniform filter reflects the typical grid resolution of wall-resolved LES and has also been reconsidered by \textcite{Bauer2019} for comparison; it is also very similar to the one used by \textcite{Piomelli1996} for channel flow. Since our focus is on the effect of the filter kernel $G$, we intentionally keep the nominal filter scale $\lambda$ constant throughout for all three filter types considered. Here, we do not report the influence of the filter width onto the energy flux and the different correlations. However, some previous testing on the effect of the filter size was performed and confirmed that the findings are not highly sensitive to the filter width.

Independent of the type of $G$ we choose, all filter operations are conducted in Fourier space. Hence, the convolution in eq.~\eqref{eq:explicitFilter} becomes the less expensive multiplication
\begin{align}
\overline{u^{\prime}_{i}}^{\lambda_{\theta}\times\lambda_{z}}\left(r,\theta,z,t\right) =
\text{FFT}^{-1}\left\{
\widehat{G}^{\lambda_{\theta}\times\lambda_{z}}\left(\kappa_{\theta},\kappa_{z}\right)\cdot
\widehat{u^{\prime}}_{i}\left(r,\kappa_{\theta},\kappa_{z},t\right)
\right\}
\text{,}
\label{eq:explicitFilterFourierSpace}
\end{align}
where
\begin{align}
\widehat{u^{\prime}}_{i} \left(r,\kappa_{\theta},\kappa_{z},t\right) =
\text{FFT}\left\{u^{\prime}_{i}\left(r,\theta,z,t\right)\right\}
\label{eq:fft}
\end{align}
are the Fourier coefficients obtained by applying a 2D fast Fourier transformation (FFT) to the velocity fluctuation field. The respective wavenumbers are denoted by $\kappa_i$, and $\widehat{G}^{\lambda_{\theta}\times \lambda_{z}}$ is the transfer function of the respective 2D filter kernel, as detailed in appendix~\ref{app:filterKernel}.

Since we filter only in $\theta$ and $z$, which are naturally periodic directions in our DNS data, the required forward and backward FFT operations are accurate and computationally very efficient. The filtering, the energy flux calculation, and the statistics are performed as post-processing on our pipe flow DNS data (section~\ref{sec:dnsDataBase}) using our python tool box \eflux, which is freely available and easily adaptable to other generic flow geometries and filter widths.

\section{Inter-scale energy fluxes based on sharp spectral scale separation}
\label{sec:resultsFourier}

We first focus on the results obtained using the sharp spectral cut-off kernel. Results for the Gaussian and box kernels are discussed later in Section~\ref{sec:filterEffect}, but the corresponding plots are presented here alongside the sharp-spectral results to enable one-to-one comparisons.

\subsection{Instantaneous inter-scale energy flux}
\label{sec:instantaneousFieldsFourier}

\begin{figure*}
\centering
\includegraphics[width=1.00\textwidth, trim=3mm 2mm 3mm 3mm, clip=true]{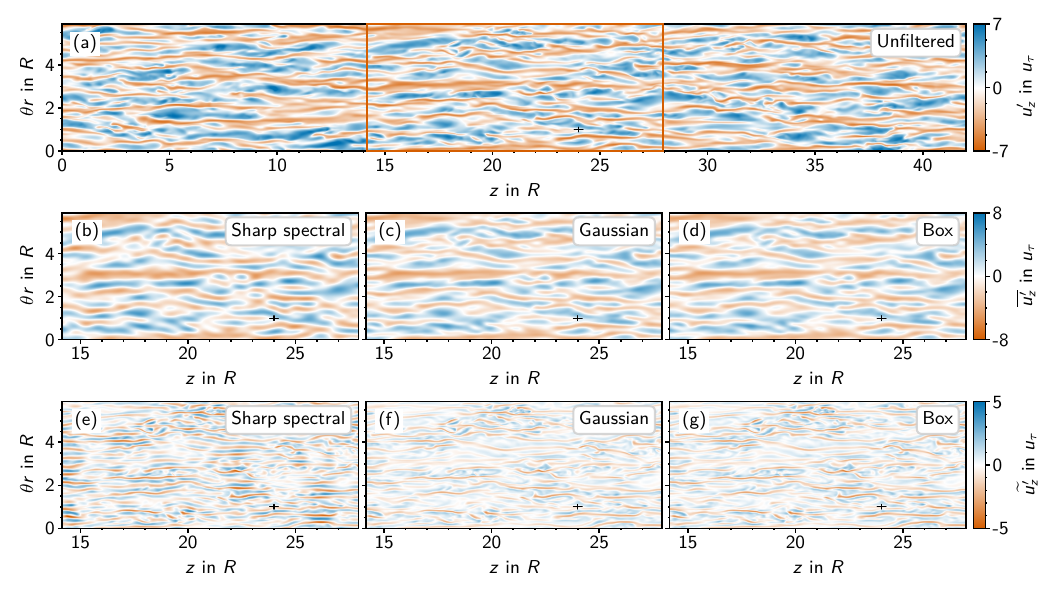}
\caption{Effect of filter kernel on the filtered velocity field in a wall-parallel ($\theta$-$z$) plane located in the buffer layer ($y^{+}=\num{12}$, $r=\num{0.93}R$). (a): Colour-coded representation of typical \highspeed ($u^{\prime}_{z}>0$) and \lowspeed ($u^{\prime}_{z}<0$) streaks in the original (unfiltered) flow field. (b), (c), (d): Coarse-grained streaks  in a small region indicated by the red box for three different kernels (sharp spectral, Gaussian, box) and one fixed nominal filter length scale $\left(\lambda^{+}_{\theta}\times\lambda^{+}_{z}\right) = \left(\num{40}\times\num{75}\right)$ as indicated by the black cross. (e), (f), (g): Removed (sub-filter) fluctuations approximately smaller than the nominal filter length scale.}
\label{fig:filteredField}
\end{figure*}

\begin{figure*}
\centering
\includegraphics[width=1.00\textwidth, trim=3mm 2mm 3mm 2mm, clip=true]{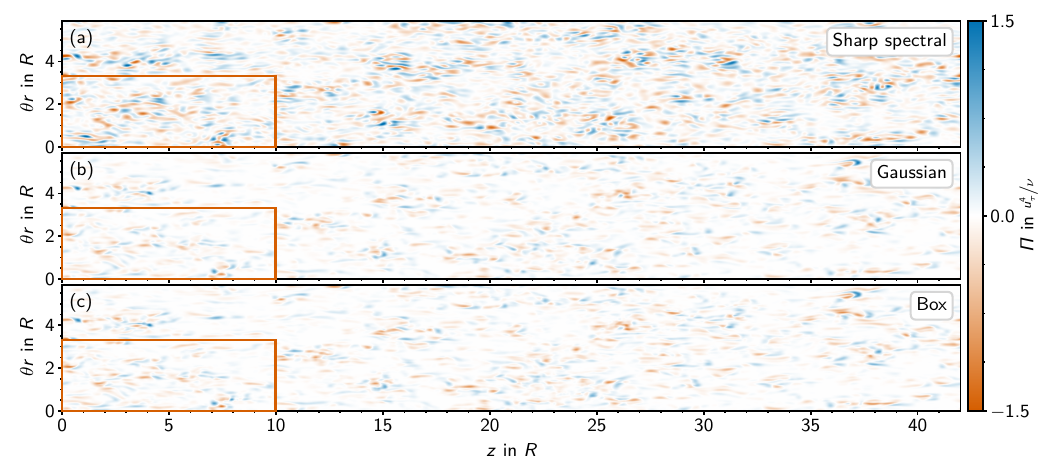}
\caption{Local structure of the inter-scale energy flux, $\Pi$, in a wall-parallel ($\theta$-$z$) plane located in the buffer layer ($y^{+}=\num{12}$, $r=\num{0.93}R$) for one arbitrary snapshot based on three different filter kernels (sharp spectral, Gaussian, box). The red box indicates the region shown in more detail in figure~\ref{fig:plotPiStreaksCompare}.}
\label{fig:plotPiFieldCompare}
\end{figure*}

The inter-scale energy flux is directly computed from the spatially filtered flow field according to eq.~\eqref{eq:eFlux}. As an example, figures~\ref{fig:filteredField}a and b show the original ($u^{\prime}_{z}$) and the coarse-grained ($\overline{u^{\prime}}_z$) streamwise velocity fluctuations in a wall-parallel plane located in the buffer layer ($y^{+}=\num{12}$, $r=\num{0.93}R$) using a 2D sharp spectral kernel for scale separation. Figure~\ref{fig:filteredField}e represents the removed (sub-filter) scales, which are approximately smaller than the chosen filter width $\left(\lambda^{+}_{\theta} \times \lambda^{+}_{z}\right) = \left(\num{40} \times \num{75} \right)$. The corresponding energy flux field ($\Pi$) shown in figure~\ref{fig:plotPiFieldCompare}a is highly intermittent in space and time.

Visual inspection of figures~\ref{fig:plotPiFieldCompare}a and
\ref{fig:filteredField}a reveals that $\Pi$ comprises smaller scales than
$u^{\prime}_{z}$ and that strong instantaneous flux events are typically much
smaller compared to streamwise streaks; at least for the given $\lambda$
considered here. This is consistent with observations reported by
\textcite{Bauer2019} and comes as no surprise in view of eq.~\eqref{eq:eFlux},
since the absolute value of the energy flux is proportional to the spatial
derivatives of the velocity field.

\begin{figure*}
\centering
\includegraphics[width=1.00\textwidth, trim=3mm 2mm 3mm 3mm, clip=true]{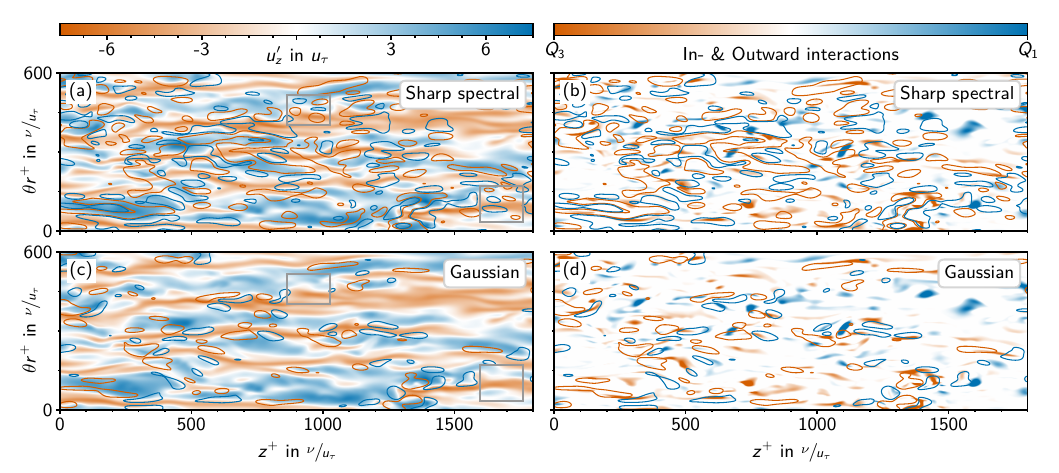}
\caption{Local structure of the inter-scale energy flux ($\Pi$) compared to typical structural features in the buffer layer ($y^{+} = 12$, $r = 0.93R$). The region shown here is indicated by the red box in figure~\ref{fig:plotPiFieldCompare} and only significant \forward (\positive) and \backward (\negative) scatter events from that figure are shown here as contour lines for $\Pi=\sfrac{\pm1}{10}\max\lvert\Pi\rvert$. (a), (b): Significant flux events based on sharp spectral filtering. (c), (d): Significant flux events based on Gaussian filtering. (a), (c): Significant flux events on top of \highspeed ($u^{\prime}_{z}>0$) and \lowspeed ($u^{\prime}_{z}<0$) streaks. \textcolor{Grey}{Grey boxes} highlight examples for distinct differences with respect to different filter kernels. (b), (d): Significant flux events on top of \inward ($Q_{3}$) and \outward ($Q_{1}$) interactions. Results for the box filter are almost identical with the ones for the Gaussian filter and therefore not shown.}
\label{fig:plotPiStreaksCompare}
\end{figure*}

Intense flux events seem to line up in pairs of alternating sign and form much large clusters (figure~\ref{fig:plotPiFieldCompare}a). From figure~\ref{fig:plotPiStreaksCompare}a  it appears that strong \backward transfer events ($\Pi<\num{0}$) more often sit on top of \highspeed streaks ($u^{\prime}_{z}>\num{0}$) and that \forward transfer events ($\Pi>\num{0}$) more often sit on top of \lowspeed streaks ($u^{\prime}_{z}<\num{0}$). For better visualisation, we choose a threshold of $\sfrac{\pm1}{10}\max\lvert\Pi\rvert$ to define strong flux events, based on the integral energy argument in \textcite{Feldmann2018}.

Both observations are perfectly consistent
with \citet{Piomelli1991} and \citet{Piomelli1996}, who reported that fluxes of
different
sign occur side-by-side and in close proximity. Based on conditional mean flow
structures, they showed that strong \backward scatter usually occurs on the
downwash side of a streamwise vortex, where fluid is transported towards the
wall, while strong \forward scatter tends to occur on its upwash side, where
fluid is transported away from the wall. From that \textcite{Piomelli1996}
inferred that \backward scatter is connected to sweep ($Q_{4}$) events and that
\forward scatter is connected to ejection ($Q_{2}$) events, which in turn are
usually associated with \high and \lowspeed streaks. In contrast to
\textcite{Piomelli1996}, we explicitly extract $Q$ events and the streamwise
vorticity ($\omega_{z}$) and compare them directly to the instantaneous flux
field. Our comparisons with $Q_{2}$, $Q_{4}$ and $\omega_{z}$ (all three not
shown here) lead basically to the same conclusions drawn from
figure~\ref{fig:plotPiStreaksCompare}a and thus further confirm and  expand the
work of \citet{Piomelli1996}. \par Figure~\ref{fig:plotPiStreaksCompare}b shows
the same instantaneous $\Pi$ events as in
figure~\ref{fig:plotPiStreaksCompare}a, but this time on top of a colour-coded
map representing localised inward ($Q_{3}$) and outward ($Q_{1}$) interactions,
which were not discussed by \textcite{Piomelli1996}. Typical energy flux events
are in general of comparable size or slightly bigger than typical $Q_{3}$ and
$Q_{1}$ events. Figure~\ref{fig:plotPiStreaksCompare}b reveals no preferred
arrangement of significant flux events with regard to \inward and \outward
interactions.

\subsection{One-point statistics}
\label{sec:onePointStatisticsFourier}

\begin{figure*}
\centering
\includegraphics[width=1.0\textwidth]{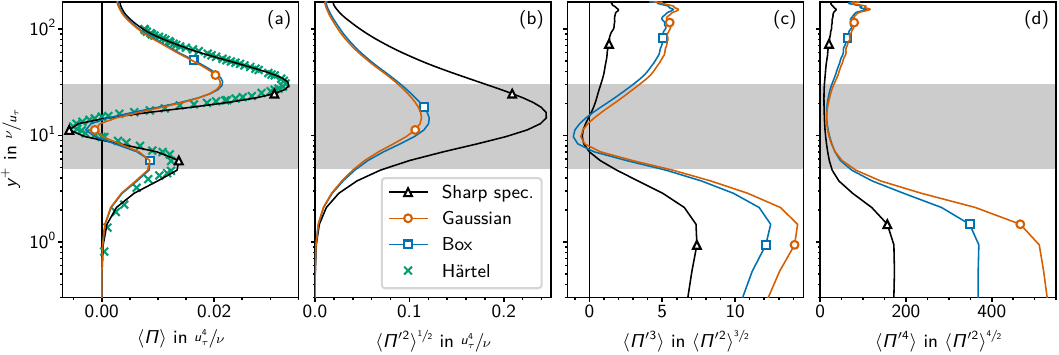}
\caption{One-point statistics for the inter-scale energy flux ($\Pi$) as a function of the wall distance ($y^+$) in viscous units ($\sfrac{\nu}{u_\tau}$) based on three different filter kernels (sharp spectral, Gaussian, box). (a): Mean flux profile, $\lla\Pi\rra$, as in eq.~\eqref{eq:reynoldsDecomposition}. (b): Root mean square profile, $\lla\Pi^{\prime2}\rra^{\sfrac{1}{2}}$, according to eq.~\eqref{eq:rms}. (c): Skewness profiles, $\lla\Pi^{\prime3}\rra/\lla\Pi^{\prime2}\rra^{\sfrac{3}{2}}$, according to eq.~\eqref{eq:skewness}. (d): Flatness profiles, $\lla\Pi^{\prime4}\rra/\lla\Pi^{\prime2}\rra^{\sfrac{4}{2}}$, according to eq.~\eqref{eq:flatness}. The grey shading marks the buffer layer ($\num{5} < y^+ < \num{30}$) and green crosses represent reference data by \textcite{Haertel1994} based on sharp spectral filtering.}
\label{fig:plotPiStatCompare}
\end{figure*}

The mean energy flux is shown in figure~\ref{fig:plotPiStatCompare}a as a function of the wall-normal distance, $y^+=\left(R-r\right)\ReTau$, scaled in inner units ($\sfrac{\nu}{u_\tau}$). Very close and far away from the wall, energy is transported from larger to smaller scales on the average ($\langle\Pi\rangle>\num{0}$); as in the classical energy cascade. In the buffer layer, on the other hand, a region of predominant negative flux becomes discernible with a net transport of energy from smaller to larger scales. This indeed indicates a localised inverse energy cascade, which roughly coincides with the region of maximal mean shear and peak streamwise turbulence intensity (figure~\ref{fig:streaksAndFlux}c) as was also observed earlier by \textcite{Haertel1994} for both, channel and pipe flow. Our $\langle\Pi\rangle$ profile compares very well with the pipe data of Härtel, who also used a sharp spectral kernel for scale separation. The net backward scatter peaks at a wall-normal distance of $y^{+}=\num{12}$; very similar to the location where the production of turbulent kinetic energy (not shown here) also reaches its maximum ($y^{+}=\num{15}$).

The root mean square (RMS) values $\langle\Pi^{\prime 2}\rangle^{\sfrac{1}{2}}$ depicted in figure~\ref{fig:plotPiStatCompare}b are in general high. This indicates a very active and intermittent instantaneous flux field with large excursions from the net energy transport $\langle\Pi\rangle$, which is orders of magnitude smaller. The ratio between the strongest flux events in the buffer layer (figure~\ref{fig:plotPiFieldCompare}) and the local mean value (figure~\ref{fig:plotPiStatCompare}a) is roughly \num{330}.

In general, near-wall turbulence is dominated by intermittent localised events: Sweeps and ejections, inward/outward interactions, and velocity spikes, just to name a few examples with decreasing probability of occurrence. In between these events -- in a spatial as well as in a temporal sense -- lie regions of relative calm. This typically leads to high kurtosis (flatness) values for the velocity field. For example, \textcite{Bauer2017} report flatness values exceeding \num{30} very close to the pipe wall due to extreme rare wall-normal velocity spike events. Note, that the flatness of a normal Gaussian distribution is three. The local flatness of $\Pi$ in the buffer layer is much higher than that and even more extreme in the outer region ($y^{+}>\num{30}$) as well as in the viscous sub-layer ($y^{+}<\num{5}$), as depicted in figure~\ref{fig:plotPiStatCompare}d. This highlights the importance of rare but extreme energy-flux events. \textcite{Piomelli1991} and \citet{Piomelli1996} also concluded that only a few, very energetic events are responsible for a large percentage of the net energy transfer between scales.

\subsection{Two-point statistics in the buffer layer}
\label{sec:twoPointStatisticsFourier}

\begin{figure*}
\centering
\includegraphics[width=1.0\textwidth]{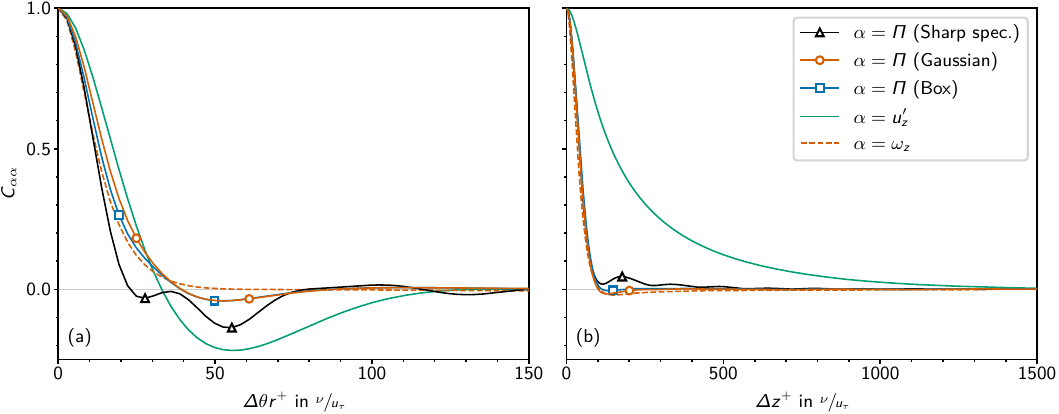}
\caption{One-dimensional two-point auto-correlations for streamwise streaks ($u^\prime_z$), quasi-streamwise vortices ($\omega_z$), and energy fluxes ($\Pi$) in the buffer layer ($y^+=12$, $r=0.93R$), based on three different filter kernels (sharp spectral, Gaussian, and box). (a): As a function of azimuthal separation ($\Delta \theta r^+$), according to eq.~\eqref{eq:twoPointStatisticsTh}. (b) As a function of axial separation ($\Delta z^+$), according to eq.~\eqref{eq:twoPointStatisticsZ}. Note that both separations are expressed as physical length scales in viscous units ($\sfrac{\nu}{u_\tau}$) and that due to symmetries, only positive separations are shown.}
\label{fig:plotPiCorr1dPiCompare}
\end{figure*}

Here we analyse the local structure of the inter-scale energy flux in comparison to different structural features derived from the near-wall velocity field in the buffer layer, where the net backward transport of energy is maximal ($y^{+}=\num{12}$). The auto-correlations of the streamwise vorticity shown in figure~\ref{fig:plotPiCorr1dPiCompare} indicate that here in this wall-parallel plane, typical streamwise vortices are roughly \num{100} viscous units wide and \num{200} viscous units long, since $C_{\omega_{z}\omega_{z}}$ approaches zero at an azimuthal separation of around $\Delta\theta r^+ = \num{50}$ and at an axial separation of around $\Delta z^+=\num{100}$. Typical streaks are roughly $\num{70}$ viscous units wide and $\num{3000}$ viscous units long at this wall-normal location. The azimuthal spacing related to the alternating nature of high- and low-speed streaks is roughly $\num{110}$ viscous units, since $C_{u^\prime_z u^\prime_z}$ is maximally anti-correlated at around $\num{55}$ viscous units azimuthal separation.

The auto-correlations for the energy flux statistically confirm our observations from the instantaneous snapshots discussed in section~\ref{sec:instantaneousFieldsFourier}. On the average, structures in the $\Pi$ field are somewhat slimmer and much shorter than typical streamwise streaks. Instead, typical $\Pi$ events are very similar in length compared to the average streamwise vortex.

\begin{figure*}
\centering
\includegraphics[width=1.0\textwidth]{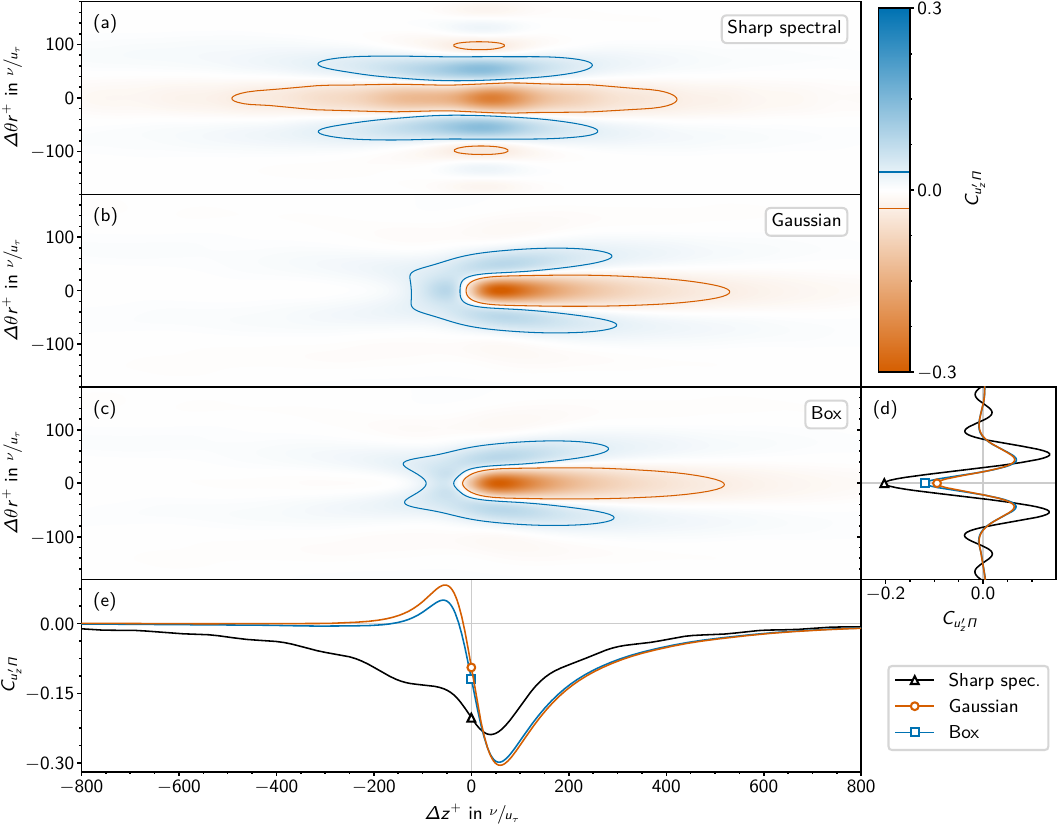}
\caption{Two-point cross-correlations between streamwise streaks ($u^{\prime}_{z}$) and local energy fluxes ($\Pi$) in a wall-parallel ($\theta$--$z$) plane located in the buffer layer ($y^{+} = 12$, $r = 0.93R$), based on three different filter kernels (sharp spectral, Gaussian, and box). (a), (b), (c): Colour-coded contour maps of the 2D correlations as a function of azimuthal ($\Delta\theta r^+$) and axial ($\Delta z^+$) separation, according to eq.~\eqref{eq:twoPointStatisticsThZ}. Note that both separations are expressed as physical length scales in viscous units ($\sfrac{\nu}{u_\tau}$). Contour lines indicate $\pm\SI{10}{\percent}$ of the absolute maximum correlation value. (d): 1D correlation profiles for azimuthal separation ($\Delta\theta r^+$), according to eq.~\eqref{eq:twoPointStatisticsTh}. (e): 1D correlation profiles for axial separation ($\Delta z^+$), according to eq.~\eqref{eq:twoPointStatisticsZ}.}
\label{fig:plotPiCorr2d1dStreaksCompare}
\end{figure*}

To further quantify our observations from section~\ref{sec:instantaneousFieldsFourier} and to extend the work of \textcite{Piomelli1996} in this regard, we compute 1D and 2D two-point cross-correlations for the energy flux with different structural features derived from the near-wall velocity field, as detailed in Appendix~\ref{app:statisticalAnalysis}. In figures~\ref{fig:plotPiCorr2d1dOmegaZCompare}a and \ref{fig:plotPiCorr2d1dStreaksCompare}a we present 2D cross-correlations with streamwise vortices ($C_{\omega_{z}\Pi}$) and streamwise streaks ($C_{u^{\prime}_{z}\Pi}$), respectively.

\begin{figure*}
\centering
\includegraphics[width=1.0\textwidth]{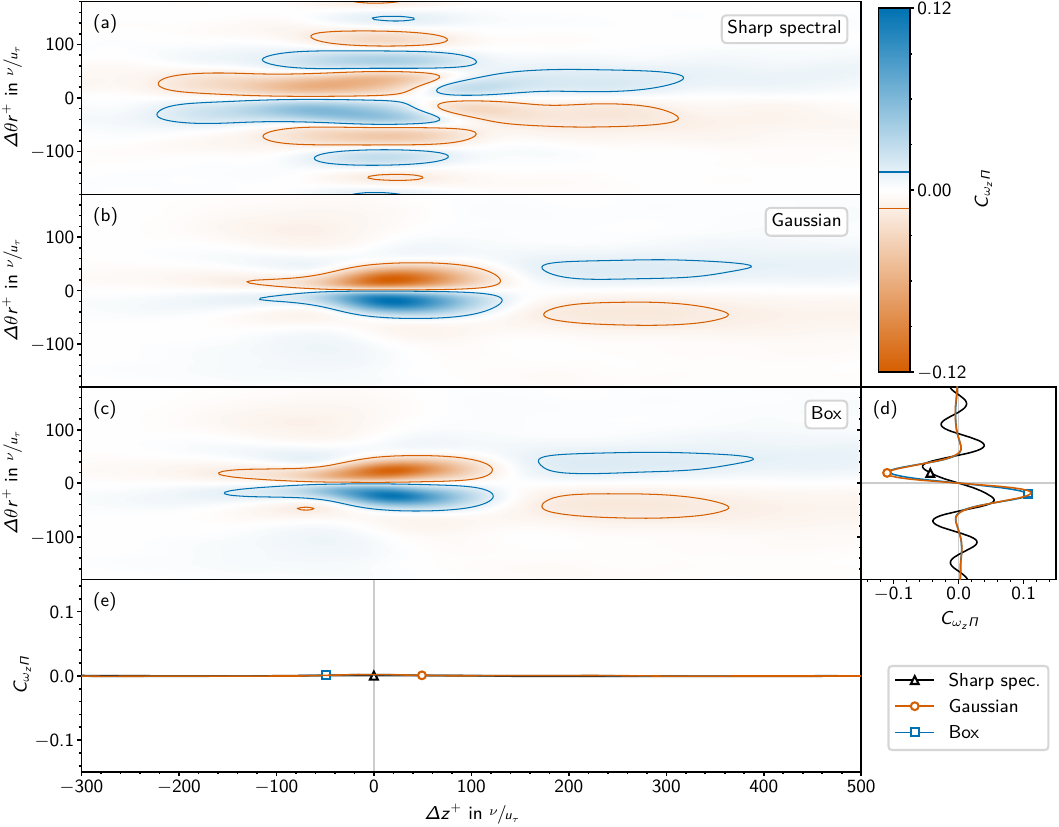}
\caption{Two-point cross-correlations between between quasi-streamwise vortices ($\omega_{z}$) and local energy fluxes ($\Pi$) in a wall-parallel ($\theta$--$z$) plane located in the buffer layer ($y^{+} = 12$, $r = 0.93R$), based on three different filter kernels (sharp spectral, Gaussian, and box). (a), (b), (c): Colour-coded contour maps of the 2D correlations as a function of azimuthal ($\Delta\theta r^+$) and axial ($\Delta z^+$) separation, according to eq.~\eqref{eq:twoPointStatisticsThZ}. Note that both separations are expressed as physical length scales in viscous units ($\sfrac{\nu}{u_\tau}$). Contour lines indicate $\pm\SI{10}{\percent}$ of the absolute maximum correlation value. (d): 1D correlation profiles for azimuthal separation ($\Delta\theta r^+$), according to eq.~\eqref{eq:twoPointStatisticsTh}. (e): 1D correlation profiles for axial separation ($\Delta z^+$), according to eq.~\eqref{eq:twoPointStatisticsZ}.}
\label{fig:plotPiCorr2d1dOmegaZCompare}
\end{figure*}

At the reference point (zero separation) $C_{\omega_{z}\Pi}$ is zero and exhibits an almost perfectly anti-symmetric behaviour in $\theta$, with maximum correlation (anti-correlation) for a negative (positive) azimuthal separation of $\num{29}$ viscous units. This means that on the average, either positive $\Pi$ events sit on the right-hand-side of a negative vortex and on the left-hand-side of a positive vortex, or that, negative $\Pi$ events sit on the right-hand-side of a positive vortex and on the left-hand-side of a negative vortex. This confirms what \textcite{Piomelli1996} inferred from their conditionally averaged flow field structures (see also section~\ref{sec:instantaneousFieldsFourier}). The 1d axial correlation is expected to be statistically zero, because of the perfect azimuthal anti-symmetry of the vortex pairs. The curves shown in figure~\ref{fig:plotPiCorr2d1dOmegaZCompare}e and their negligible departure from zero highlight the quality of the statistical sample used to compute the two-point correlations presented in this study.

At the reference point, $C_{u^{\prime}_{z}\Pi}$ is in general \negative (Fig.~\ref{fig:plotPiCorr2d1dStreaksCompare}a). This indicates, that on the average \backward transfer events ($\Pi<\num{0}$) more often sit on top of \highspeed streaks ($u^{\prime}_{z}>\num{0}$) and that \forward transfer events ($\Pi>\num{0}$) more often sit on top of \lowspeed streaks ($u^{\prime}_{z}<\num{0}$); further confirming our observations in section~\ref{sec:instantaneousFieldsFourier} and the conclusions of \textcite{Piomelli1996}. The cross-correlations $C_{Q_{4}\Pi}$ and $C_{Q_{2}\Pi}$ were also computed, but basically support \textcite{Piomelli1996} in the same way and are therefore not shown. The correlations between $u^{\prime}_{z}$ and $\Pi$ are largely streamwise-symmetric for axial separations ($\Delta z$), as shown in figures~\ref{fig:plotPiCorr2d1dStreaksCompare}a and e. This means that there is no preferred upstream-downstream orientation of energy flux events with regard to \positive and \negative streaks, when the sharp spectral kernel is used for scale separation.

\begin{figure*}
\centering
\includegraphics[width=1.0\textwidth]{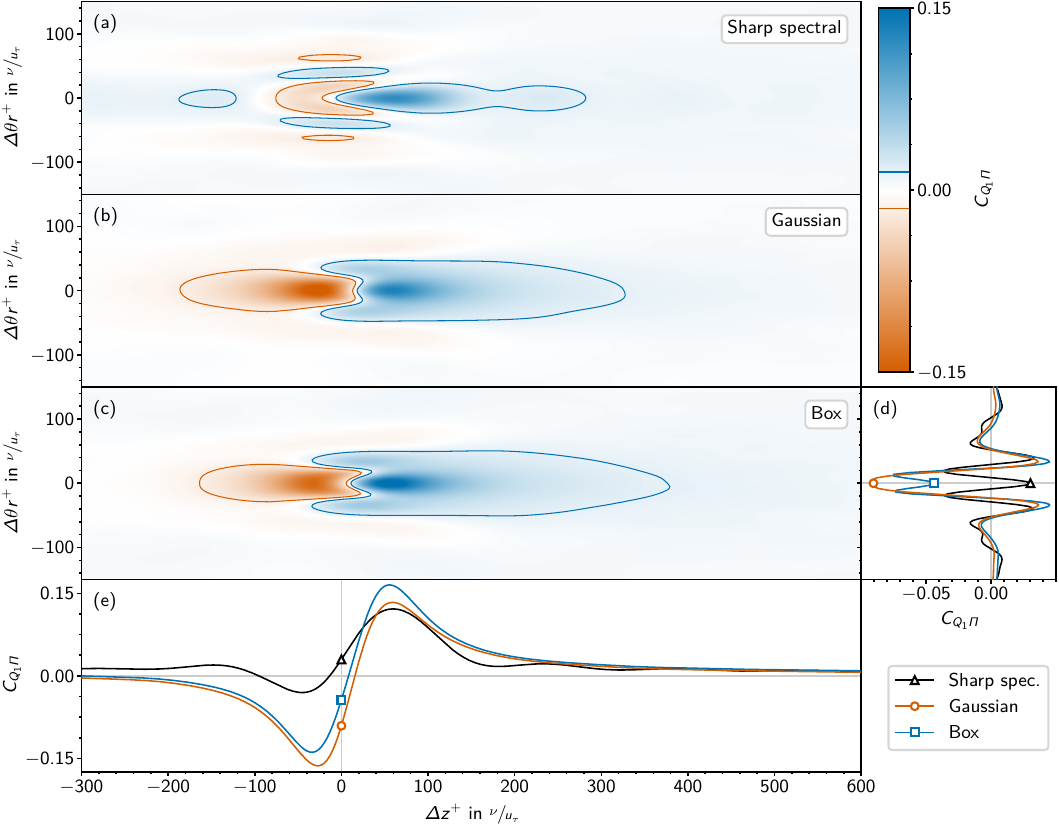}
\caption{Two-point cross-correlations between outward interactions ($Q_{1}$) and local energy fluxes ($\Pi$) in a wall-parallel ($\theta$--$z$) plane located in the buffer layer ($y^{+} = 12$, $r = 0.93R$), based on three different filter kernels (sharp spectral, Gaussian, and box). (a), (b), (c): Colour-coded contour maps of the 2D correlations as a function of azimuthal ($\Delta\theta r^+$) and axial ($\Delta z^+$) separation, according to eq.~\eqref{eq:twoPointStatisticsThZ}. Note that both separations are expressed as physical length scales in viscous units ($\sfrac{\nu}{u_\tau}$). Contour lines indicate $\pm\SI{10}{\percent}$ of the absolute maximum correlation value. (d): 1D correlation profiles for azimuthal separation ($\Delta\theta r^+$), according to eq.~\eqref{eq:twoPointStatisticsTh}. (e): 1D correlation profiles for axial separation ($\Delta z^+$), according to eq.~\eqref{eq:twoPointStatisticsZ}.}
\label{fig:plotPiCorr2d1dQ1Compare}
\end{figure*}

The cross-correlation $C_{Q_{1}\Pi}$ is shown in figure~\ref{fig:plotPiCorr2d1dQ1Compare}a and demonstrates the structural connection between typical outward interactions and the direction of the energy cascade in the buffer layer. On the average, fluid transport away from the wall via outward interactions ($Q_1<\num{0}$) coincides with \backward scatter ($\Pi<\num{0}$), since $C_{Q_{1}\Pi}>\num{0}$ for zero separation, when the sharp spectral kernel is used for scale separation. A weak negative correlation is confined to a short region around $\Delta z^{+}=\num{-50}$, meaning that localised \forward scatter events seem to sit preferably shortly upstream of an $Q_1$ event. For all other axial shifts ($\Delta z^{+}$), the correlation is positive and a pronounced upstream-downstream asymmetry can be distinguished; in contrast to other cross-correlations based on sharp spectral filtering (\eg figure~\ref{fig:plotPiCorr2d1dStreaksCompare}). This suggests, that patterns of localised \backward-\forward-\backward scatter preferably occur simultaneously with outward interactions, where the most downstream \backward scatter event overlaps the $Q_{1}$ event with its tail. Vice versa, triples of \forward-\backward-\forward scatter seem to occur in connection with fluid transport towards the wall via inward interactions, where the central \backward event overlaps the $Q_{3}$ event with its head (not shown).

Additionally, figure~\ref{fig:plotPiCorr1dPiCompare}a reveals that the energy flux field based on the sharp spectral kernel features several azimuthal oscillations suggesting a uniform stacking of $\Pi$ events similar to the alternating nature of high- and low-speed streaks: Regions of positive energy flux are frequently flanked by regions of weaker negative flux and vice versa. This behaviour is amplified in connection with the different structural features in the buffer layer leading to strong azimuthal oscillations in all cross-correlations, which clearly resemble the $\operatorname{sinc}$ function shape of the sharp spectral kernel in physical space, see eq.~\eqref{eq:kernelFourierPhys}, as for example shown in panels (a) and (d) of figures.~\ref{fig:plotPiCorr2d1dOmegaZCompare} to \ref{fig:plotPiCorr2d1dQ1Compare}.

\section{Effect of different filter kernels on the inter-scale energy flux}
\label{sec:filterEffect}

We now extend our discussion to include the Gaussian and box kernel options, focusing on how the choice of kernel affects the local energy flux field and its statistics.

\subsection{Instantaneous energy flux}
\label{sec:instantaneousFieldsCompare}

A strong effect of the choice of filter kernel can readily be observed in the sub-filter part of the velocity field presented in figures~\ref{fig:filteredField}e to g. In the coarse-grained velocity snapshots, the differences are more subtle to spot (Figs.~\ref{fig:filteredField}b to d). Already in the filtered velocity fields subtle differences in dependence on the kernel type are observed. The coarse-grained field based on the sharp spectral kernel shows a tendency to form patterns of alternating high and low speed areas. Thus, elongated high- and low-speed areas that appear inclined (spreading across $\theta_r$) are converted to more speckled areas with less incline. It is important to note, that these checkerboard-type effects of the sharp spectral kernel do occur independently of the filter width but become more obvious for larger filter widths and are absent for the Gaussian and the box filter kernels.

Since $\Pi$ contains spatial derivatives and second order terms of the filtered velocity, it is not surprising, that differences due to the filter kernel manifest themselves even stronger in the local structure of $\Pi$. Figure~\ref{fig:plotPiFieldCompare} compares the instantaneous flux field in the buffer layer for all three kernels considered here. For the Gaussian and the box filter, the energy flux field appears less speckled and less active in terms of population and intensity, when compared to the sharp spectral filter. In some regions, where the sharp spectral filter seems to generate very symmetric square patterns, the Gaussian and the box filter generate no significant events at all, as for example highlighted by the grey boxes shown in figures~\ref{fig:plotPiStreaksCompare}a and c. Moreover, for the sharp spectral kernel the spatial structure of strong $\Pi$ events appears to be slightly shifted with respect to both other kernels, what becomes more obvious when using \eg streaks or $Q_1$ events as visual frame of reference (figures~\ref{fig:plotPiStreaksCompare}). Although not explicitly
noted, the same discrepancies can be observed in the instantaneous $\Pi$ fields presented in \textcite{Buzzicotti2018}, who consider the effect of different filter kernels on the scale energetics in HIT. This shift in the energy flux events can be traced back to the differences in the filtered velocity field and the speckles of checkerboard structure of the sharp spectral kernel.

Both the Gaussian and the box kernels yield energy fluxes with somewhat smaller and more equally sized events. For the sharp spectral kernel, some of the largest \backscatter events even span multiple streaks in $\theta$ direction. Moreover, for the Gaussian and the box filters, significant $\Pi$ events appear to align with the shear layers between streaks. Most of the time, the orientation of the major axis of the elliptical flux events is thereby slightly tilted against the direction of the mean flow ($z$) alongside the azimuthal meander of the streaks; another observation absent for the sharp spectral case. The picture for the box kernel is almost identical to the one for the Gaussian kernel (\cf figures~\ref{fig:filteredField} and \ref{fig:plotPiFieldCompare}) and therefore not explicitly shown here.

The strong energy flux events based on the Gaussian and the box filter seem to accumulate at locations of highly active inward ($Q_{3}$) and outward ($Q_{1}$) interactions, while for the sharp spectral based $\Pi$ field, this effect is much less pronounced (figures~\ref{fig:plotPiStreaksCompare}b and d). Similar observations can be made when comparing the different kernels with respect to sweep ($Q_{4}$) and ejection ($Q_{2}$) events (not shown here). Thus, the attributed role of localised $Q$ events for the local energy flux seems to be a different one depending on the type of filter kernel used for scale separation at the considered filter width.

\subsection{One-point statistics}
\label{sec:onePointStatisticsCompare}

Qualitatively, all filter kernels produce similar mean statistics. Regions of positive and negative net flux are consistent among all filters and the inner ($y^{+}=\num{5}$) and outer ($y^{+}=\num{30}$) maxima of forward scatter collapse exactly with the boundaries of the buffer layer. Also the backscatter maxima collapse for all three filters; namely at $y^{+}=\num{12}$. Similar qualitative agreements among different kernels can be observed for the other statistical moments (RMS, skewness and flatness), which are plotted in figures~\ref{fig:plotPiStatCompare}b to d. However, quantitatively the sharp spectral kernel yields much stronger mean fluxes compared to Gaussian and box. Forward peak fluxes are up to \SI{55}{\percent} higher and the backscatter peak continuously reduces from sharp spectral to box to Gaussian (figure~\ref{fig:plotPiStatCompare}a). This is consistent with \textcite{Leslie1979}, who showed analytically that the amount of backward scatter contribution to the eddy viscosity is significantly reduced for a Gaussian filter in comparison to a sharp spectral filter. Results from a preliminary study\cite{Chen2019} indicate that small variations in $\lambda$ are insufficient to align the peaks of the mean energy flux profiles obtained with the Gaussian kernel with those obtained using the sharp spectral kernel. Since minor adjustments to the filter scale fail to produce consistent mean energy flux statistics across different kernels, we conclude that the influence of the filter kernel cannot, in general, be compensated by simply changing $\lambda$.

Additionally, the RMS statistics presented in figure~\ref{fig:plotPiStatCompare}b reveal twice as strong excursions from the mean flux throughout the buffer layer, when comparing the sharp spectral filter against Gaussian and box. These effects were already inferred from the instantaneous picture (figure~\ref{fig:plotPiFieldCompare}) and are consistent with \textcite{Aoyama2005} and \textcite{Piomelli1991}, who report a much smaller volume fraction of backscatter events in HIT and channel flow, when comparing Gaussian against sharp spectral filters.

The skewness and flatness factors shown in figures~\ref{fig:plotPiStatCompare}c and d, exhibit discrepancies up to \SI{300}{\percent}, when comparing sharp spectral, Gaussian and box. The observed discrepancies arise from the known speckling effect introduced by the sharp spectral kernel. This effect generates localised, artificial $\Pi$ events in physical space that are comparatively strong and spatially coherent, which can overshadow genuine intermittent structures in the $\Pi$ field. As a result, rare physical events appear statistically less extreme, leading to a reduction in flatness. Moreover, the speckle-induced structures are symmetric in nature, which diminishes the inherent asymmetry of the physical $\Pi$ field. This causes the overall skewness to be reduced when using the sharp spectral filter, in contrast to the Gaussian or box filters that preserve the field's asymmetry more faithfully. Similar deviations have recently been reported by \textcite{Cardesa2015}, \textcite{Buzzicotti2018}, and \textcite{Cardesa2019}, who considered the effect of different filters and different inter-scale energy transfer markers in HIT: The sharp spectral filter leads to much more symmetric (\ie less skewed) probability density functions compared to the Gauss filter. 

As already discussed in section~\ref{sec:onePointStatisticsFourier}, rare but extreme $\Pi$ events play an important role for reliable prediction of scale energetics. Especially in the viscous sublayer and throughout the entire outer region, the sharp spectral filter predicts significantly fewer and more uniformly distributed extreme events as the Gaussian and box kernels do.

Both the Gaussian and box filters yield nearly identical one-point statistics throughout most of the pipe domain. The only notable differences occur in the skewness and flatness factors in the immediate vicinity of the wall, within the viscous sublayer. Currently, we have no robust explanation for the origin of this discrepancy.

\subsection{Two-point statistics in the buffer layer}
\label{sec:twoPointStatisticsCompare}

For Gaussian and box kernels, typical energy flux structures are roughly \num{75} viscous units wide and \num{200} viscous units long, and thus \SI{67}{\percent} thicker (figure~\ref{fig:plotPiCorr1dPiCompare}a) but \SI{58}{\percent} shorter (figure~\ref{fig:plotPiCorr1dPiCompare}b) than in the sharp spectral case. In particular, the sharp spectral kernel generates $\Pi$ structures which are statistically longer than typical $\omega_{z}$ structures, while the other two filters predict flux events of the same length as typical streamwise vortices. Additionally, the sharp spectral kernel generates inter-scale energy fluxes with two dominant azimuthal anti-correlations, which are not present for the Gaussian and the box kernels (figure~\ref{fig:plotPiCorr1dPiCompare}a). This fact suggests, that the oscillatory behaviour of the flux field based on sharp spectral scale separation is unphysical and rather related to artificial ringing effects \cite{Gonzalez2002, Mishra2014} which is also causes the checkerboard-type pattern in the filtered velocity fields.

For all three filter kernels, the correlation between streaks and energy fluxes displays the same sign at zero separation ($C_{u^{\prime}_{z}\Pi}< \num{0}$). However, the Gaussian and the box filters predict a roughly \SI{50}{\percent} lower anti-correlation at the reference point, which is best seen in figures~\ref{fig:plotPiCorr2d1dStreaksCompare}d and e. In contrast to the streamwise symmetric behaviour of the sharp spectral filter (section~\ref{sec:twoPointStatisticsFourier}), Gaussian and box reveal asymmetric correlations for streamwise separations, as shown in figures~\ref{fig:plotPiCorr2d1dStreaksCompare}b, c and e. This indicates that \backward scatter events ($\Pi<\num{0}$) are more often located on the upstream edges of a \highspeed streak and on the downstream edges of a \lowspeed streak, since a downstream shift ($\Delta z>\num{0}$) does not change the negative sign of the correlation factor for shifts of at least $\Delta z^{+}\approx\num{800}$, while an upstream shift ($\Delta z^{+}<\num{0}$) leads to a weak positive correlation factor after a mean shift of only $\Delta z^{+}\approx\num{-40}$. Similarly, \forward scatter events ($\Pi>\num{0}$) are more often located on the
upstream edges of \lowspeed streaks ($u^{\prime}_{z}$<\num{0}).

It is important to note that, the length/width of the structures matters here, since a shift of a pair of the generally shorter \forward and \backward scatter events over an elongated structure of unique sign cannot cause an asymmetric correlation function. Bearing this in mind, the highly streamwise-symmetric appearance of the correlation for the sharp spectral kernel signifies that energy flux events are mostly centred with regard to the streamwise streaks. Contrarily, in case of the Gaussian and the box kernels the cross-correlations suggest that energy flux events are located on the inclined meander of the streaks where \highspeed and \lowspeed regions meet, as already anticipated from the instantaneous snapshots (figure~\ref{fig:plotPiStreaksCompare}).

The cross-correlations with outward interactions ($C_{Q_{1}\Pi}$) are negative at the reference point for the Gaussian and the box kernels (figures~\ref{fig:plotPiCorr2d1dQ1Compare}b and c). This is an important contrast to the positive correlation observed for the sharp spectral kernel. It implies that fluid transport towards the outer region of the flow is associated with upscale energy transfer when the sharp spectral kernel is used for scale separation (section~\ref{sec:twoPointStatisticsFourier}), whereas for the Gauss and the box kernels the opposite is observed. Energy is transferred to smaller scales at locations where fluid is leaving the wall via outward interactions ($Q_1$). However, all three kernels equally predict that instantaneous backscatter ($\Pi<\num{0}$) usually occurs directly downstream of an outward interaction with the peak positive correlation at around $\Delta z^+ \approx\num{150}$ axial separation (figure~\ref{fig:plotPiCorr2d1dQ1Compare}).

Further, for small downstream shifts up to $\Delta z^+\approx\num{30}$ and for all upstream shifts, the correlations with $Q_{1}$ events remain negative for the Gaussian and the box kernels (figures~\ref{fig:plotPiCorr2d1dQ1Compare}b and c). This is in contrast to the cross-correlations for the sharp spectral filter and indicates that for Gaussian and box $Q_{1}$ events more likely occur together with \forward-\backward patterns instead of with \backward-\forward-\backward triplets (figure~\ref{fig:plotPiCorr2d1dQ1Compare}b). Similar differences between filter kernels can be observed for cross-correlations with inward interactions ($Q_3$), which are not shown here.

For separations in $\theta$, auto-correlations based on the Gaussian and the box kernels do not exhibit azimuthal oscillations in the energy flux field, as seen for the sharp spectral kernel (figure~\ref{fig:plotPiCorr1dPiCompare}). Equally, none of the cross-correlations based on the Gaussian and the box kernels show an oscillatory behaviour as strong as seen for the sharp spectral kernel. Instead, they feature only one pair of strong correlation peaks on each side of the reference point, before they decay to zero (\eg figures~\ref{fig:plotPiCorr2d1dOmegaZCompare}, \ref{fig:plotPiCorr2d1dStreaksCompare} and \ref{fig:plotPiCorr2d1dQ1Compare}). This analysis complies with the observation of sparser distributed instantaneous $\Pi$ events for the Gaussian and the box kernel and more densely populated $\Pi$ events in case of the sharp spectral kernel (figures~\ref{fig:plotPiFieldCompare}a, b, c) and it also matches the interpretation of figure~\ref{fig:plotPiCorr2d1dStreaksCompare}e. If strong $\Pi$ events are associated with the strong shear layers between the streaks (Gaussian and box) rather than with streaks themselves (sharp spectral), then there is less space available where the condition for this correlation is provided.

\section{Discussion}
\label{sec:discussion}

Increasingly often the transfer of kinetic energy across scales based on spatial low-pass filtering is analysed in order to better understand the energy budget of turbulent flows in connection with typical structures appearing in the flow field \cite[\eg][]{Cardesa2019, Kelley2013, Piomelli1996, Haertel1994}. Of particular interest is the potential role of the inverse energy cascade for the formation of large-scale structures in the turbulent flow field, which is known to occur in 2d flows \cite[\eg][]{vonKameke2011}, in 3d flows with suppressed 3d movement \cite[\eg][]{Xia2011} as well as in wall-bounded flows \cite{Haertel1994}. In this study, we analysed how different filter kernels affect the instantaneous energy flux field, its statistics, and its interplay with typical structures in the buffer layer of a turbulent pipe flow. Our quantitative assessment brought to light the following insights:

\begin{itemize}

\item The instantaneous picture of the energy flux largely depends on the shape of the filter type used for scale separation: For the sharp spectral kernel, it is more speckled, more active, and slightly shifted when compared to Gauss and box at the considered filter width. In general, significant energy flux events appear to be spatially localised. When a sharp spectral filter kernel is used for scale separation they appear, however, longer (in $z$) but thinner (in $\theta$) than in the case of the other to kernels. These observations are supported by one-point statistics for all wall-normal distances. For the mean energy flux the findings are not sensitive to the filter widths as has been confirmed in previous tests. Differences with regard to the filter kernel also appear in two-point auto- and cross-correlations at one particular location in the buffer layer ($y^+=\num{12}$). Size and exact spatial alignment of $\Pi$ events is of major importance when analysing its role for the dynamics, mixing and transport properties of wall-bounded turbulent flows. For example, \textcite{Kelley2013} related energy flux interfaces (\ie its zero lines) to hyperbolic Lagrangian coherent structures in order to separate regions in a 2d turbulent flow depending on their dynamical transport properties.

\item Analysing the shape of the two-point cross-correlations in the buffer layer revealed further differences regarding the local structure and the alignment of the inter-scale energy fluxes, when different filters of the same filter widths are used. For example, cross-correlations with streaks are largely streamwise symmetric for the sharp spectral filter kernel, whereas they are highly asymmetric for Gauss and box, reflecting the anisotropic nature of the dynamics of the wall-cycle. For the Gauss and the box filter, backward scatter occurs preferably on an upstream edge of a high-speed streak, whereas forward scatter occurs preferably on an upstream edge of a low-speed streak at the considered filter width. Also, for these two filters relevant $\Pi$ events preferably sit on the inclined meander at the borders of a streak. When the sharp spectral filter kernel is used for scale separation by contrast, relevant $\Pi$ events sit centred on top of the streaks and cross-correlations with streamwise vortices are weaker and more dispersed. In this case, strong flux events would be explained by the mere strength and size of a streak, whereas, when the Gauss or the box filter is used to compute $\Pi$, the meandering large-scale azimuthal instability of the streaks seems to be responsible for strong local flux events. In that case, the high local fluxes might also be interpreted as an energy source for the meandering motion and a trigger for the streak instability. 

\item For the considered filter width cross-correlations with outward interactions have different signs at the reference point and therefore indicate a contradicting structural connection with strong inter-scale flux events. Fluid transport away from the wall directly coincides with backward scatter when the sharp spectral filter kernel is used, whereas for Gauss and box, energy is transferred to smaller scales at locations where fluid is leaving the wall via outward interactions ($Q_{1}$). In the light of recent findings about the inverse energy cascade and the spiralling behaviour of the scale-energy paths that start from the buffer layer and diverge, feeding longer and wider turbulent structures reaching into the outer regions of the flow \cite{Cimarelli2013, Cimarelli2016}, these differences matter. In particular, the picture drawn by the Gauss and the box filter gives only little support for any direct association of net backscatter with flow structures ascending from the buffer layer at the considered filter width, whereas the sharp spectral filter kernel filter does indeed support \textcite{Cimarelli2013} and \citet{Cimarelli2016} in this regard. However, all three kernels equally predict that instantaneous backscatter frequently occurs directly downstream of an outward interaction.

\item One-point statistics of the energy flux are qualitatively similar for the sharp spectral, Gauss and box kernel, and thus robust with respect to the filter type for all wall-normal locations. Our study confirms the net inverse energy cascade in the buffer layer reported by \textcite{Haertel1994} by demonstrating its existence for all filter kernels used here. Regions of positive and negative net energy transfer are consistent among filters and the peak flux locations all collapse at $y^+=\num{5}$, $y^+=\num{12}$ and $y^+=\num{30}$, respectively.

\item However, changing the filter kernel from Gauss or box to sharp spectral generates quantitatively very different one-point statistics that are not sensitive to the filter width as was tested previously \cite{Chen2019}. Our study revealed, that the prediction of strong flux events in the buffer layer and rare but extreme events outside of the buffer layer is very sensitive to the type of filter chosen for scale separation. For example, the sharp spectral kernel predicts more than twice as high net energy transfer rates in the buffer layer and deviates by a factor of up to three in the higher order moments in the outer region and the viscous sublayer. In the same way, the choice of the filter type might affect interpretations regarding superstructures when turning to higher Reynolds number flows and larger filter widths \cite[\eg][]{Bauer2019}. In this regard it is important to note that, the outer region hosts extreme large structures (\eg VLSM in case of pipe flow) with very slow dynamics (\ie rare events) \cite{Wu2012} and that their footprints play an important role for drag generation in the viscous sublayer \cite{Oerlue2011}. 

\end{itemize}

To conclude, we observed that the diagnostic tool $\Pi$ is highly sensitive to the type of filter implemented for scale separation at the considered filter width. We unravelled the complicated response of $\Pi$ to different filters and found astonishing qualitative agreement for the one-point statistics at all wall-normal locations, although there are large deviations in the intensity and distribution of localised flux events. The sharp spectral kernel is known to act non-local in physical space and therefore generates artificial ringing in physical space \cite{Gonzalez2002, Mishra2014}. Our analysis indicates that this behaviour of the sharp spectral filter is mainly responsible for the major deviations discussed above. Additionally, the subgrid scale kinetic energy based on sharp spectral kernels is not guaranteed to be positive\cite{Vreman1994}, whereas positivity of the SGS stresses when using positive kernels (\eg Gauss and box) was shown more recently using a simple concavity argument\cite{Sadek2018}. Therefore, we suggest that the sharp spectral filter should be used with caution when $\Pi$ is used as an analysis tool to investigate the scale-energetics in combination with spatially localised structures in wall-bounded turbulent flows.

\begin{acknowledgments}
This work was funded by the German Research Foundation (DFG) through the Priority Programme \turbspp. Alexandra von Kameke gratefully acknowledges DFG support through grant \texttt{KA 4854/1-1}. Mohammad Umair acknowledges the Erasmus${}^+$ scholarship and the stipend provided by the University of Bremen during his stay at ZARM. Computational resources were provided by \hlrn through project \texttt{hbi00041}, which is also gratefully acknowledged. We thank Jan Chen for implementing an initial version of \eflux and for evaluating the effect of filter size on the energy flux in the context of his Master's thesis \cite{Chen2019}. Finally, we are grateful to the anonymous referees for their constructive feedback and many helpful suggestions.
\end{acknowledgments}

The data that support the findings of this study are openly available in \pangaea.

\bibliography{pof2025.bib}

\begin{thebibliography}{63}%
\makeatletter
\providecommand \@ifxundefined [1]{%
 \@ifx{#1\undefined}
}%
\providecommand \@ifnum [1]{%
 \ifnum #1\expandafter \@firstoftwo
 \else \expandafter \@secondoftwo
 \fi
}%
\providecommand \@ifx [1]{%
 \ifx #1\expandafter \@firstoftwo
 \else \expandafter \@secondoftwo
 \fi
}%
\providecommand \natexlab [1]{#1}%
\providecommand \enquote  [1]{``#1''}%
\providecommand \bibnamefont  [1]{#1}%
\providecommand \bibfnamefont [1]{#1}%
\providecommand \citenamefont [1]{#1}%
\providecommand \href@noop [0]{\@secondoftwo}%
\providecommand \href [0]{\begingroup \@sanitize@url \@href}%
\providecommand \@href[1]{\@@startlink{#1}\@@href}%
\providecommand \@@href[1]{\endgroup#1\@@endlink}%
\providecommand \@sanitize@url [0]{\catcode `\\12\catcode `\$12\catcode
  `\&12\catcode `\#12\catcode `\^12\catcode `\_12\catcode `\%12\relax}%
\providecommand \@@startlink[1]{}%
\providecommand \@@endlink[0]{}%
\providecommand \url  [0]{\begingroup\@sanitize@url \@url }%
\providecommand \@url [1]{\endgroup\@href {#1}{\urlprefix }}%
\providecommand \urlprefix  [0]{URL }%
\providecommand \Eprint [0]{\href }%
\providecommand \doibase [0]{http://dx.doi.org/}%
\providecommand \selectlanguage [0]{\@gobble}%
\providecommand \bibinfo  [0]{\@secondoftwo}%
\providecommand \bibfield  [0]{\@secondoftwo}%
\providecommand \translation [1]{[#1]}%
\providecommand \BibitemOpen [0]{}%
\providecommand \bibitemStop [0]{}%
\providecommand \bibitemNoStop [0]{.\EOS\space}%
\providecommand \EOS [0]{\spacefactor3000\relax}%
\providecommand \BibitemShut  [1]{\csname bibitem#1\endcsname}%
\let\auto@bib@innerbib\@empty
\bibitem [{\citenamefont {Leonard}(1975)}]{Leonard1975}%
  \BibitemOpen
  \bibfield  {author} {\bibinfo {author} {\bibfnamefont {A.}~\bibnamefont
  {Leonard}},\ }\bibfield  {title} {{\selectlanguage {English}\enquote
  {\bibinfo {title} {{Energy Cascade in Large-Eddy Simulations of Turbulent
  Fluid Flows}},}\ }\ }(\bibinfo  {publisher} {Elsevier},\ \bibinfo {year}
  {1975})\ pp.\ \bibinfo {pages} {237--248}\BibitemShut {NoStop}%
\bibitem [{\citenamefont {Germano}(1992)}]{Germano1992}%
  \BibitemOpen
  \bibfield  {author} {\bibinfo {author} {\bibfnamefont {M.}~\bibnamefont
  {Germano}},\ }\bibfield  {title} {\enquote {\bibinfo {title} {{Turbulence:
  the filtering approach}},}\ }\href {\doibase 10.1017/S0022112092001733}
  {\bibfield  {journal} {\bibinfo  {journal} {Journal of Fluid Mechanics}\
  }\textbf {\bibinfo {volume} {238}},\ \bibinfo {pages} {325--336} (\bibinfo
  {year} {1992})}\BibitemShut {NoStop}%
\bibitem [{\citenamefont {Liu}, \citenamefont {Meneveau},\ and\ \citenamefont
  {Katz}(1994)}]{Liu1994}%
  \BibitemOpen
  \bibfield  {author} {\bibinfo {author} {\bibfnamefont {S.}~\bibnamefont
  {Liu}}, \bibinfo {author} {\bibfnamefont {C.}~\bibnamefont {Meneveau}}, \
  and\ \bibinfo {author} {\bibfnamefont {J.}~\bibnamefont {Katz}},\ }\bibfield
  {title} {\enquote {\bibinfo {title} {{On the properties of similarity
  subgrid-scale models as deduced from measurements in a turbulent jet}},}\
  }\href {\doibase 10.1017/S0022112094002296} {\bibfield  {journal} {\bibinfo
  {journal} {Journal of Fluid Mechanics}\ }\textbf {\bibinfo {volume} {275}},\
  \bibinfo {pages} {83--119} (\bibinfo {year} {1994})}\BibitemShut {NoStop}%
\bibitem [{\citenamefont {Tao}, \citenamefont {Katz},\ and\ \citenamefont
  {Meneveau}(2002)}]{Tao2002}%
  \BibitemOpen
  \bibfield  {author} {\bibinfo {author} {\bibfnamefont {B.}~\bibnamefont
  {Tao}}, \bibinfo {author} {\bibfnamefont {J.}~\bibnamefont {Katz}}, \ and\
  \bibinfo {author} {\bibfnamefont {C.}~\bibnamefont {Meneveau}},\ }\bibfield
  {title} {\enquote {\bibinfo {title} {{Statistical geometry of subgrid-scale
  stresses determined from holographic particle image velocimetry
  measurements}},}\ }\href {\doibase 10.1017/S0022112001007443} {\bibfield
  {journal} {\bibinfo  {journal} {Journal of Fluid Mechanics}\ }\textbf
  {\bibinfo {volume} {457}},\ \bibinfo {pages} {35--78} (\bibinfo {year}
  {2002})}\BibitemShut {NoStop}%
\bibitem [{\citenamefont {Bai}, \citenamefont {Meneveau},\ and\ \citenamefont
  {Katz}(2013)}]{Bai2013}%
  \BibitemOpen
  \bibfield  {author} {\bibinfo {author} {\bibfnamefont {K.}~\bibnamefont
  {Bai}}, \bibinfo {author} {\bibfnamefont {C.}~\bibnamefont {Meneveau}}, \
  and\ \bibinfo {author} {\bibfnamefont {J.}~\bibnamefont {Katz}},\ }\bibfield
  {title} {\enquote {\bibinfo {title} {{Experimental study of spectral energy
  fluxes in turbulence generated by a fractal, tree-like object}},}\ }\href
  {\doibase 10.1063/1.4819351} {\bibfield  {journal} {\bibinfo  {journal}
  {Physics of Fluids}\ }\textbf {\bibinfo {volume} {25}} (\bibinfo {year}
  {2013}),\ 10.1063/1.4819351}\BibitemShut {NoStop}%
\bibitem [{\citenamefont {Drivas}\ \emph {et~al.}(2017)\citenamefont {Drivas},
  \citenamefont {Johnson}, \citenamefont {Lalescu},\ and\ \citenamefont
  {Wilczek}}]{Drivas2017}%
  \BibitemOpen
  \bibfield  {author} {\bibinfo {author} {\bibfnamefont {T.~D.}\ \bibnamefont
  {Drivas}}, \bibinfo {author} {\bibfnamefont {P.~L.}\ \bibnamefont {Johnson}},
  \bibinfo {author} {\bibfnamefont {C.~C.}\ \bibnamefont {Lalescu}}, \ and\
  \bibinfo {author} {\bibfnamefont {M.}~\bibnamefont {Wilczek}},\ }\bibfield
  {title} {\enquote {\bibinfo {title} {{Large-scale sweeping of small-scale
  eddies in turbulence: A filtering approach}},}\ }\href {\doibase
  10.1103/PhysRevFluids.2.104603} {\bibfield  {journal} {\bibinfo  {journal}
  {Physical Review Fluids}\ }\textbf {\bibinfo {volume} {2}},\ \bibinfo {pages}
  {1--16} (\bibinfo {year} {2017})}\BibitemShut {NoStop}%
\bibitem [{\citenamefont {Bian}\ and\ \citenamefont {Aluie}(2019)}]{Bian2019}%
  \BibitemOpen
  \bibfield  {author} {\bibinfo {author} {\bibfnamefont {X.}~\bibnamefont
  {Bian}}\ and\ \bibinfo {author} {\bibfnamefont {H.}~\bibnamefont {Aluie}},\
  }\bibfield  {title} {\enquote {\bibinfo {title} {{Decoupled Cascades of
  Kinetic and Magnetic Energy in Magnetohydrodynamic Turbulence}},}\ }\href
  {\doibase 10.1103/PhysRevLett.122.135101} {\bibfield  {journal} {\bibinfo
  {journal} {Physical Review Letters}\ }\textbf {\bibinfo {volume} {122}},\
  \bibinfo {pages} {135101} (\bibinfo {year} {2019})}\BibitemShut {NoStop}%
\bibitem [{\citenamefont {Araki}, \citenamefont {Vela-Mart{\'{i}}n},\ and\
  \citenamefont {Lozano-Dur{\'{a}}n}(2024)}]{Araki2024}%
  \BibitemOpen
  \bibfield  {author} {\bibinfo {author} {\bibfnamefont {R.}~\bibnamefont
  {Araki}}, \bibinfo {author} {\bibfnamefont {A.}~\bibnamefont
  {Vela-Mart{\'{i}}n}}, \ and\ \bibinfo {author} {\bibfnamefont
  {A.}~\bibnamefont {Lozano-Dur{\'{a}}n}},\ }\bibfield  {title} {\enquote
  {\bibinfo {title} {{Forgetfulness of turbulent energy cascade associated with
  different mechanisms}},}\ }\href {\doibase 10.1088/1742-6596/2753/1/012001}
  {\bibfield  {journal} {\bibinfo  {journal} {Journal of Physics: Conference
  Series}\ }\textbf {\bibinfo {volume} {2753}},\ \bibinfo {pages} {012001}
  (\bibinfo {year} {2024})}\BibitemShut {NoStop}%
\bibitem [{\citenamefont {Zhou}, \citenamefont {Ding},\ and\ \citenamefont
  {Cheng}(2024)}]{Zhou2024}%
  \BibitemOpen
  \bibfield  {author} {\bibinfo {author} {\bibfnamefont {Z.}~\bibnamefont
  {Zhou}}, \bibinfo {author} {\bibfnamefont {J.}~\bibnamefont {Ding}}, \ and\
  \bibinfo {author} {\bibfnamefont {W.}~\bibnamefont {Cheng}},\ }\bibfield
  {title} {\enquote {\bibinfo {title} {{Mixing and inter-scale energy transfer
  in Richtmyer–Meshkov turbulence}},}\ }\href {\doibase 10.1017/jfm.2024.240}
  {\bibfield  {journal} {\bibinfo  {journal} {Journal of Fluid Mechanics}\
  }\textbf {\bibinfo {volume} {984}},\ \bibinfo {pages} {A56} (\bibinfo {year}
  {2024})}\BibitemShut {NoStop}%
\bibitem [{\citenamefont {Park}\ and\ \citenamefont
  {Lozano-Dur{\'{a}}n}(2025)}]{Park2025}%
  \BibitemOpen
  \bibfield  {author} {\bibinfo {author} {\bibfnamefont {D.}~\bibnamefont
  {Park}}\ and\ \bibinfo {author} {\bibfnamefont {A.}~\bibnamefont
  {Lozano-Dur{\'{a}}n}},\ }\bibfield  {title} {\enquote {\bibinfo {title} {{The
  coherent structure of the energy cascade in isotropic turbulence}},}\ }\href
  {\doibase 10.1038/s41598-024-80698-3} {\bibfield  {journal} {\bibinfo
  {journal} {Scientific Reports}\ }\textbf {\bibinfo {volume} {15}},\ \bibinfo
  {pages} {14} (\bibinfo {year} {2025})}\BibitemShut {NoStop}%
\bibitem [{\citenamefont {Piomelli}, \citenamefont {Yu},\ and\ \citenamefont
  {Adrian}(1996)}]{Piomelli1996}%
  \BibitemOpen
  \bibfield  {author} {\bibinfo {author} {\bibfnamefont {U.}~\bibnamefont
  {Piomelli}}, \bibinfo {author} {\bibfnamefont {Y.}~\bibnamefont {Yu}}, \ and\
  \bibinfo {author} {\bibfnamefont {R.~J.}\ \bibnamefont {Adrian}},\ }\bibfield
   {title} {\enquote {\bibinfo {title} {{Subgrid-scale energy transfer and
  near-wall turbulence structure}},}\ }\href {\doibase 10.1063/1.868829}
  {\bibfield  {journal} {\bibinfo  {journal} {Physics of Fluids}\ }\textbf
  {\bibinfo {volume} {8}},\ \bibinfo {pages} {215--224} (\bibinfo {year}
  {1996})}\BibitemShut {NoStop}%
\bibitem [{\citenamefont {Miller}, \citenamefont {Estejab},\ and\ \citenamefont
  {Bailey}(2014)}]{Miller2014}%
  \BibitemOpen
  \bibfield  {author} {\bibinfo {author} {\bibfnamefont {M.~A.}\ \bibnamefont
  {Miller}}, \bibinfo {author} {\bibfnamefont {B.}~\bibnamefont {Estejab}}, \
  and\ \bibinfo {author} {\bibfnamefont {S.~C.~C.}\ \bibnamefont {Bailey}},\
  }\bibfield  {title} {\enquote {\bibinfo {title} {{Evaluation of hot-wire
  spatial filtering corrections for wall turbulence and correction for
  end-conduction effects}},}\ }\href {\doibase 10.1007/s00348-014-1735-3}
  {\bibfield  {journal} {\bibinfo  {journal} {Experiments in Fluids}\ }\textbf
  {\bibinfo {volume} {55}},\ \bibinfo {pages} {1735} (\bibinfo {year}
  {2014})}\BibitemShut {NoStop}%
\bibitem [{\citenamefont {Buzzicotti}\ \emph
  {et~al.}(2018{\natexlab{a}})\citenamefont {Buzzicotti}, \citenamefont
  {Aluie}, \citenamefont {Biferale},\ and\ \citenamefont
  {Linkmann}}]{Buzzicotti2018a}%
  \BibitemOpen
  \bibfield  {author} {\bibinfo {author} {\bibfnamefont {M.}~\bibnamefont
  {Buzzicotti}}, \bibinfo {author} {\bibfnamefont {H.}~\bibnamefont {Aluie}},
  \bibinfo {author} {\bibfnamefont {L.}~\bibnamefont {Biferale}}, \ and\
  \bibinfo {author} {\bibfnamefont {M.}~\bibnamefont {Linkmann}},\ }\bibfield
  {title} {\enquote {\bibinfo {title} {{Energy transfer in turbulence under
  rotation}},}\ }\href {\doibase 10.1103/PhysRevFluids.3.034802} {\bibfield
  {journal} {\bibinfo  {journal} {Physical Review Fluids}\ }\textbf {\bibinfo
  {volume} {3}},\ \bibinfo {pages} {034802} (\bibinfo {year}
  {2018}{\natexlab{a}})}\BibitemShut {NoStop}%
\bibitem [{\citenamefont {Bauer}, \citenamefont {von Kameke},\ and\
  \citenamefont {Wagner}(2019)}]{Bauer2019}%
  \BibitemOpen
  \bibfield  {author} {\bibinfo {author} {\bibfnamefont {C.}~\bibnamefont
  {Bauer}}, \bibinfo {author} {\bibfnamefont {A.}~\bibnamefont {von Kameke}}, \
  and\ \bibinfo {author} {\bibfnamefont {C.}~\bibnamefont {Wagner}},\
  }\bibfield  {title} {\enquote {\bibinfo {title} {{Kinetic energy budget of
  the largest scales in turbulent pipe flow}},}\ }\href {\doibase
  10.1103/PhysRevFluids.4.064607} {\bibfield  {journal} {\bibinfo  {journal}
  {Physical Review Fluids}\ }\textbf {\bibinfo {volume} {4}},\ \bibinfo {pages}
  {064607} (\bibinfo {year} {2019})}\BibitemShut {NoStop}%
\bibitem [{\citenamefont {Dogan}\ \emph {et~al.}(2019)\citenamefont {Dogan},
  \citenamefont {{\"{O}}rl{\"{u}}}, \citenamefont {Gatti}, \citenamefont
  {Vinuesa},\ and\ \citenamefont {Schlatter}}]{Dogan2019}%
  \BibitemOpen
  \bibfield  {author} {\bibinfo {author} {\bibfnamefont {E.}~\bibnamefont
  {Dogan}}, \bibinfo {author} {\bibfnamefont {R.}~\bibnamefont
  {{\"{O}}rl{\"{u}}}}, \bibinfo {author} {\bibfnamefont {D.}~\bibnamefont
  {Gatti}}, \bibinfo {author} {\bibfnamefont {R.}~\bibnamefont {Vinuesa}}, \
  and\ \bibinfo {author} {\bibfnamefont {P.}~\bibnamefont {Schlatter}},\
  }\bibfield  {title} {\enquote {\bibinfo {title} {{Quantification of amplitude
  modulation in wall-bounded turbulence}},}\ }\href {\doibase
  10.1088/1873-7005/aaca81} {\bibfield  {journal} {\bibinfo  {journal} {Fluid
  Dynamics Research}\ }\textbf {\bibinfo {volume} {51}} (\bibinfo {year}
  {2019}),\ 10.1088/1873-7005/aaca81}\BibitemShut {NoStop}%
\bibitem [{\citenamefont {Kawata}\ and\ \citenamefont
  {Tsukahara}(2021)}]{Kawata2021}%
  \BibitemOpen
  \bibfield  {author} {\bibinfo {author} {\bibfnamefont {T.}~\bibnamefont
  {Kawata}}\ and\ \bibinfo {author} {\bibfnamefont {T.}~\bibnamefont
  {Tsukahara}},\ }\bibfield  {title} {\enquote {\bibinfo {title} {{Scale
  interactions in turbulent plane Couette flows in minimal domains}},}\ }\href
  {\doibase 10.1017/jfm.2020.1063} {\bibfield  {journal} {\bibinfo  {journal}
  {Journal of Fluid Mechanics}\ }\textbf {\bibinfo {volume} {911}},\ \bibinfo
  {pages} {A55} (\bibinfo {year} {2021})}\BibitemShut {NoStop}%
\bibitem [{\citenamefont {Kannadasan}, \citenamefont {Atkinson},\ and\
  \citenamefont {Soria}(2024)}]{Kannadasan2024}%
  \BibitemOpen
  \bibfield  {author} {\bibinfo {author} {\bibfnamefont {E.}~\bibnamefont
  {Kannadasan}}, \bibinfo {author} {\bibfnamefont {C.}~\bibnamefont
  {Atkinson}}, \ and\ \bibinfo {author} {\bibfnamefont {J.}~\bibnamefont
  {Soria}},\ }\bibfield  {title} {\enquote {\bibinfo {title} {{Interplay of
  scales during the spatial evolution of energy-containing motions in
  wall-bounded turbulent flows}},}\ }\href {\doibase 10.1017/jfm.2024.267}
  {\bibfield  {journal} {\bibinfo  {journal} {Journal of Fluid Mechanics}\
  }\textbf {\bibinfo {volume} {985}},\ \bibinfo {pages} {A6} (\bibinfo {year}
  {2024})}\BibitemShut {NoStop}%
\bibitem [{\citenamefont {Eyink}(1994)}]{Eyink1994}%
  \BibitemOpen
  \bibfield  {author} {\bibinfo {author} {\bibfnamefont {G.~L.}\ \bibnamefont
  {Eyink}},\ }\bibfield  {title} {\enquote {\bibinfo {title} {{Energy
  dissipation without viscosity in ideal hydrodynamics I. Fourier analysis and
  local energy transfer}},}\ }\href {\doibase 10.1016/0167-2789(94)90117-1}
  {\bibfield  {journal} {\bibinfo  {journal} {Physica D: Nonlinear Phenomena}\
  }\textbf {\bibinfo {volume} {78}},\ \bibinfo {pages} {222--240} (\bibinfo
  {year} {1994})}\BibitemShut {NoStop}%
\bibitem [{\citenamefont {Eyink}(1995)}]{Eyink1995}%
  \BibitemOpen
  \bibfield  {author} {\bibinfo {author} {\bibfnamefont {G.~L.}\ \bibnamefont
  {Eyink}},\ }\bibfield  {title} {\enquote {\bibinfo {title} {{Local energy
  flux and the refined similarity hypothesis}},}\ }\href {\doibase
  10.1007/BF02183352} {\bibfield  {journal} {\bibinfo  {journal} {Journal of
  Statistical Physics}\ }\textbf {\bibinfo {volume} {78}},\ \bibinfo {pages}
  {335--351} (\bibinfo {year} {1995})}\BibitemShut {NoStop}%
\bibitem [{\citenamefont {Mishra}\ \emph {et~al.}(2014)\citenamefont {Mishra},
  \citenamefont {Liu}, \citenamefont {Skote},\ and\ \citenamefont
  {Fu}}]{Mishra2014}%
  \BibitemOpen
  \bibfield  {author} {\bibinfo {author} {\bibfnamefont {M.}~\bibnamefont
  {Mishra}}, \bibinfo {author} {\bibfnamefont {X.}~\bibnamefont {Liu}},
  \bibinfo {author} {\bibfnamefont {M.}~\bibnamefont {Skote}}, \ and\ \bibinfo
  {author} {\bibfnamefont {C.~W.}\ \bibnamefont {Fu}},\ }\bibfield  {title}
  {\enquote {\bibinfo {title} {{Kolmogorov spectrum consistent optimization for
  multi-scale flow decomposition}},}\ }\href {\doibase 10.1063/1.4871106}
  {\bibfield  {journal} {\bibinfo  {journal} {Physics of Fluids}\ }\textbf
  {\bibinfo {volume} {26}} (\bibinfo {year} {2014}),\
  10.1063/1.4871106}\BibitemShut {NoStop}%
\bibitem [{\citenamefont {Smits}, \citenamefont {McKeon},\ and\ \citenamefont
  {Marusic}(2011)}]{Smits2011}%
  \BibitemOpen
  \bibfield  {author} {\bibinfo {author} {\bibfnamefont {A.~J.}\ \bibnamefont
  {Smits}}, \bibinfo {author} {\bibfnamefont {B.~J.}\ \bibnamefont {McKeon}}, \
  and\ \bibinfo {author} {\bibfnamefont {I.}~\bibnamefont {Marusic}},\
  }\bibfield  {title} {\enquote {\bibinfo {title} {{High-Reynolds number wall
  turbulence}},}\ }\href {\doibase 10.1146/annurev-fluid-122109-160753}
  {\bibfield  {journal} {\bibinfo  {journal} {Annual Review of Fluid
  Mechanics}\ }\textbf {\bibinfo {volume} {43}},\ \bibinfo {pages} {353--375}
  (\bibinfo {year} {2011})}\BibitemShut {NoStop}%
\bibitem [{\citenamefont {Mizuno}(2016)}]{Mizuno2016}%
  \BibitemOpen
  \bibfield  {author} {\bibinfo {author} {\bibfnamefont {Y.}~\bibnamefont
  {Mizuno}},\ }\bibfield  {title} {\enquote {\bibinfo {title} {{Spectra of
  energy transport in turbulent channel flows for moderate Reynolds
  numbers}},}\ }\href {\doibase 10.1017/jfm.2016.564} {\bibfield  {journal}
  {\bibinfo  {journal} {Journal of Fluid Mechanics}\ }\textbf {\bibinfo
  {volume} {805}},\ \bibinfo {pages} {171--187} (\bibinfo {year}
  {2016})}\BibitemShut {NoStop}%
\bibitem [{\citenamefont {Ahn}\ and\ \citenamefont {Sung}(2017)}]{Ahn2017}%
  \BibitemOpen
  \bibfield  {author} {\bibinfo {author} {\bibfnamefont {J.}~\bibnamefont
  {Ahn}}\ and\ \bibinfo {author} {\bibfnamefont {H.~J.}\ \bibnamefont {Sung}},\
  }\bibfield  {title} {\enquote {\bibinfo {title} {{Relationship between
  streamwise and azimuthal length scales in a turbulent pipe flow}},}\ }\href
  {\doibase 10.1063/1.4997346} {\bibfield  {journal} {\bibinfo  {journal}
  {Physics of Fluids}\ }\textbf {\bibinfo {volume} {29}} (\bibinfo {year}
  {2017}),\ 10.1063/1.4997346}\BibitemShut {NoStop}%
\bibitem [{\citenamefont {Lee}\ and\ \citenamefont {Moser}(2019)}]{Lee2019}%
  \BibitemOpen
  \bibfield  {author} {\bibinfo {author} {\bibfnamefont {M.}~\bibnamefont
  {Lee}}\ and\ \bibinfo {author} {\bibfnamefont {R.~D.}\ \bibnamefont
  {Moser}},\ }\bibfield  {title} {\enquote {\bibinfo {title} {{Spectral
  analysis of the budget equation in turbulent channel flows at high Reynolds
  number}},}\ }\href {\doibase 10.1017/jfm.2018.903} {\bibfield  {journal}
  {\bibinfo  {journal} {Journal of Fluid Mechanics}\ }\textbf {\bibinfo
  {volume} {860}},\ \bibinfo {pages} {886--938} (\bibinfo {year}
  {2019})}\BibitemShut {NoStop}%
\bibitem [{\citenamefont {Cardesa}\ and\ \citenamefont
  {Lozano-Dur{\'{a}}n}(2019)}]{Cardesa2019}%
  \BibitemOpen
  \bibfield  {author} {\bibinfo {author} {\bibfnamefont {J.~I.}\ \bibnamefont
  {Cardesa}}\ and\ \bibinfo {author} {\bibfnamefont {A.}~\bibnamefont
  {Lozano-Dur{\'{a}}n}},\ }\href {\doibase
  https://doi.org/10.48550/arXiv.1912.11143} {\enquote {\bibinfo {title}
  {{Inter-scale energy transfer in turbulence from the viewpoint of subfilter
  scales}},}\ }\bibinfo {type} {Tech. Rep.}\ \bibinfo {number} {December}\
  (\bibinfo {year} {2019})\BibitemShut {NoStop}%
\bibitem [{\citenamefont {Cardesa}, \citenamefont {Vela-Mart{\'{i}}n},\ and\
  \citenamefont {Jim{\'{e}}nez}(2017)}]{Cardesa2017}%
  \BibitemOpen
  \bibfield  {author} {\bibinfo {author} {\bibfnamefont {J.~I.}\ \bibnamefont
  {Cardesa}}, \bibinfo {author} {\bibfnamefont {A.}~\bibnamefont
  {Vela-Mart{\'{i}}n}}, \ and\ \bibinfo {author} {\bibfnamefont
  {J.}~\bibnamefont {Jim{\'{e}}nez}},\ }\bibfield  {title} {\enquote {\bibinfo
  {title} {{The turbulent cascade in five dimensions}},}\ }\href {\doibase
  10.1126/science.aan7933} {\bibfield  {journal} {\bibinfo  {journal}
  {Science}\ }\textbf {\bibinfo {volume} {357}},\ \bibinfo {pages} {782--784}
  (\bibinfo {year} {2017})}\BibitemShut {NoStop}%
\bibitem [{\citenamefont {Richardson}(1922)}]{Richardson1922}%
  \BibitemOpen
  \bibfield  {author} {\bibinfo {author} {\bibfnamefont {L.~F.}\ \bibnamefont
  {Richardson}},\ }\href@noop {} {\emph {\bibinfo {title} {{Weather prediction
  by numerical process}}}}\ (\bibinfo  {publisher} {Cambridge University
  Press},\ \bibinfo {year} {1922})\BibitemShut {NoStop}%
\bibitem [{\citenamefont {Kolmogorov}(1941)}]{Kolmogorov1941}%
  \BibitemOpen
  \bibfield  {author} {\bibinfo {author} {\bibfnamefont {A.~N.}\ \bibnamefont
  {Kolmogorov}},\ }\bibfield  {title} {\enquote {\bibinfo {title} {{The local
  structure of turbulence in incompressible viscous fluid for very large
  Reynolds}},}\ }\href@noop {} {\bibfield  {journal} {\bibinfo  {journal}
  {Dokl. Akad. Nauk SSSR}\ }\textbf {\bibinfo {volume} {30}},\ \bibinfo {pages}
  {301} (\bibinfo {year} {1941})}\BibitemShut {NoStop}%
\bibitem [{\citenamefont {Eyink}\ and\ \citenamefont
  {Aluie}(2009)}]{Eyink2009}%
  \BibitemOpen
  \bibfield  {author} {\bibinfo {author} {\bibfnamefont {G.~L.}\ \bibnamefont
  {Eyink}}\ and\ \bibinfo {author} {\bibfnamefont {H.}~\bibnamefont {Aluie}},\
  }\bibfield  {title} {\enquote {\bibinfo {title} {{Localness of energy cascade
  in hydrodynamic turbulence. I. smooth coarse graining}},}\ }\href {\doibase
  10.1063/1.3266883} {\bibfield  {journal} {\bibinfo  {journal} {Physics of
  Fluids}\ }\textbf {\bibinfo {volume} {21}},\ \bibinfo {pages} {1--9}
  (\bibinfo {year} {2009})}\BibitemShut {NoStop}%
\bibitem [{\citenamefont {Aluie}\ and\ \citenamefont
  {Eyink}(2009)}]{Aluie2009}%
  \BibitemOpen
  \bibfield  {author} {\bibinfo {author} {\bibfnamefont {H.}~\bibnamefont
  {Aluie}}\ and\ \bibinfo {author} {\bibfnamefont {G.~L.}\ \bibnamefont
  {Eyink}},\ }\bibfield  {title} {{\selectlanguage {English}\enquote {\bibinfo
  {title} {{Localness of energy cascade in hydrodynamic turbulence. II. Sharp
  spectral filter}},}\ }}\href {\doibase 10.1063/1.3266948} {\bibfield
  {journal} {\bibinfo  {journal} {Physics of Fluids}\ }\textbf {\bibinfo
  {volume} {21}},\ \bibinfo {pages} {115108} (\bibinfo {year}
  {2009})}\BibitemShut {NoStop}%
\bibitem [{\citenamefont {Eyink}(2005)}]{Eyink2005}%
  \BibitemOpen
  \bibfield  {author} {\bibinfo {author} {\bibfnamefont {G.~L.}\ \bibnamefont
  {Eyink}},\ }\bibfield  {title} {\enquote {\bibinfo {title} {{Locality of
  turbulent cascades}},}\ }\href {\doibase 10.1016/j.physd.2005.05.018}
  {\bibfield  {journal} {\bibinfo  {journal} {Physica D: Nonlinear Phenomena}\
  }\textbf {\bibinfo {volume} {207}},\ \bibinfo {pages} {91--116} (\bibinfo
  {year} {2005})}\BibitemShut {NoStop}%
\bibitem [{\citenamefont {Cardesa}\ \emph {et~al.}(2015)\citenamefont
  {Cardesa}, \citenamefont {Vela-Mart{\'{i}}n}, \citenamefont {Dong},\ and\
  \citenamefont {Jim{\'{e}}nez}}]{Cardesa2015}%
  \BibitemOpen
  \bibfield  {author} {\bibinfo {author} {\bibfnamefont {J.~I.}\ \bibnamefont
  {Cardesa}}, \bibinfo {author} {\bibfnamefont {A.}~\bibnamefont
  {Vela-Mart{\'{i}}n}}, \bibinfo {author} {\bibfnamefont {S.}~\bibnamefont
  {Dong}}, \ and\ \bibinfo {author} {\bibfnamefont {J.}~\bibnamefont
  {Jim{\'{e}}nez}},\ }\bibfield  {title} {\enquote {\bibinfo {title} {{The
  temporal evolution of the energy flux across scales in homogeneous
  turbulence}},}\ }\href {\doibase 10.1063/1.4935812} {\bibfield  {journal}
  {\bibinfo  {journal} {Physics of Fluids}\ }\textbf {\bibinfo {volume} {27}},\
  \bibinfo {pages} {111702} (\bibinfo {year} {2015})}\BibitemShut {NoStop}%
\bibitem [{\citenamefont {Kline}\ \emph {et~al.}(1967)\citenamefont {Kline},
  \citenamefont {Reynolds}, \citenamefont {Schraub},\ and\ \citenamefont
  {Runstadler}}]{Kline1967}%
  \BibitemOpen
  \bibfield  {author} {\bibinfo {author} {\bibfnamefont {S.~J.}\ \bibnamefont
  {Kline}}, \bibinfo {author} {\bibfnamefont {W.~C.}\ \bibnamefont {Reynolds}},
  \bibinfo {author} {\bibfnamefont {F.~A.}\ \bibnamefont {Schraub}}, \ and\
  \bibinfo {author} {\bibfnamefont {P.~W.}\ \bibnamefont {Runstadler}},\
  }\bibfield  {title} {{\selectlanguage {English}\enquote {\bibinfo {title}
  {{The structure of turbulent boundary layers}},}\ }}\href {\doibase
  10.1017/S0022112067001740} {\bibfield  {journal} {\bibinfo  {journal}
  {Journal of Fluid Mechanics}\ }\textbf {\bibinfo {volume} {30}},\ \bibinfo
  {pages} {741--773} (\bibinfo {year} {1967})}\BibitemShut {NoStop}%
\bibitem [{\citenamefont {Jim{\'{e}}nez}\ and\ \citenamefont
  {Pinelli}(1999)}]{Jimenez1999}%
  \BibitemOpen
  \bibfield  {author} {\bibinfo {author} {\bibfnamefont {J.}~\bibnamefont
  {Jim{\'{e}}nez}}\ and\ \bibinfo {author} {\bibfnamefont {A.}~\bibnamefont
  {Pinelli}},\ }\bibfield  {title} {\enquote {\bibinfo {title} {{The autonomous
  cycle of near-wall turbulence}},}\ }\href {\doibase
  10.1017/S0022112099005066} {\bibfield  {journal} {\bibinfo  {journal}
  {Journal of Fluid Mechanics}\ }\textbf {\bibinfo {volume} {389}},\ \bibinfo
  {pages} {335--359} (\bibinfo {year} {1999})}\BibitemShut {NoStop}%
\bibitem [{\citenamefont {Wallace}(2016)}]{Wallace2016}%
  \BibitemOpen
  \bibfield  {author} {\bibinfo {author} {\bibfnamefont {J.~M.}\ \bibnamefont
  {Wallace}},\ }\bibfield  {title} {{\selectlanguage {English}\enquote
  {\bibinfo {title} {{Quadrant Analysis in Turbulence Research: History and
  Evolution}},}\ }}\href {\doibase 10.1146/annurev-fluid-122414-034550}
  {\bibfield  {journal} {\bibinfo  {journal} {Annual Review of Fluid
  Mechanics}\ }\textbf {\bibinfo {volume} {48}},\ \bibinfo {pages} {131--158}
  (\bibinfo {year} {2016})}\BibitemShut {NoStop}%
\bibitem [{\citenamefont {Marati}, \citenamefont {Casciola},\ and\
  \citenamefont {Piva}(2004)}]{Marati2004}%
  \BibitemOpen
  \bibfield  {author} {\bibinfo {author} {\bibfnamefont {N.}~\bibnamefont
  {Marati}}, \bibinfo {author} {\bibfnamefont {C.~M.}\ \bibnamefont
  {Casciola}}, \ and\ \bibinfo {author} {\bibfnamefont {R.}~\bibnamefont
  {Piva}},\ }\bibfield  {title} {\enquote {\bibinfo {title} {{Energy cascade
  and spatial fluxes in wall turbulence}},}\ }\href {\doibase
  10.1017/S0022112004001818} {\bibfield  {journal} {\bibinfo  {journal}
  {Journal of Fluid Mechanics}\ }\textbf {\bibinfo {volume} {521}},\ \bibinfo
  {pages} {191--215} (\bibinfo {year} {2004})}\BibitemShut {NoStop}%
\bibitem [{\citenamefont {Saikrishnan}\ \emph {et~al.}(2012)\citenamefont
  {Saikrishnan}, \citenamefont {De~Angelis}, \citenamefont {Longmire},
  \citenamefont {Marusic}, \citenamefont {Casciola},\ and\ \citenamefont
  {Piva}}]{Saikrishnan2012}%
  \BibitemOpen
  \bibfield  {author} {\bibinfo {author} {\bibfnamefont {N.}~\bibnamefont
  {Saikrishnan}}, \bibinfo {author} {\bibfnamefont {E.}~\bibnamefont
  {De~Angelis}}, \bibinfo {author} {\bibfnamefont {E.~K.}\ \bibnamefont
  {Longmire}}, \bibinfo {author} {\bibfnamefont {I.}~\bibnamefont {Marusic}},
  \bibinfo {author} {\bibfnamefont {C.~M.}\ \bibnamefont {Casciola}}, \ and\
  \bibinfo {author} {\bibfnamefont {R.}~\bibnamefont {Piva}},\ }\bibfield
  {title} {\enquote {\bibinfo {title} {{Reynolds number effects on scale energy
  balance in wall turbulence}},}\ }\href {\doibase 10.1063/1.3673609}
  {\bibfield  {journal} {\bibinfo  {journal} {Physics of Fluids}\ }\textbf
  {\bibinfo {volume} {24}} (\bibinfo {year} {2012}),\
  10.1063/1.3673609}\BibitemShut {NoStop}%
\bibitem [{\citenamefont {Cimarelli}, \citenamefont {De~Angelis},\ and\
  \citenamefont {Casciola}(2013)}]{Cimarelli2013}%
  \BibitemOpen
  \bibfield  {author} {\bibinfo {author} {\bibfnamefont {A.}~\bibnamefont
  {Cimarelli}}, \bibinfo {author} {\bibfnamefont {E.}~\bibnamefont
  {De~Angelis}}, \ and\ \bibinfo {author} {\bibfnamefont {C.~M.}\ \bibnamefont
  {Casciola}},\ }\bibfield  {title} {\enquote {\bibinfo {title} {{Paths of
  energy in turbulent channel flows}},}\ }\href {\doibase 10.1017/jfm.2012.528}
  {\bibfield  {journal} {\bibinfo  {journal} {Journal of Fluid Mechanics}\
  }\textbf {\bibinfo {volume} {715}},\ \bibinfo {pages} {436--451} (\bibinfo
  {year} {2013})}\BibitemShut {NoStop}%
\bibitem [{\citenamefont {Cimarelli}\ \emph {et~al.}(2016)\citenamefont
  {Cimarelli}, \citenamefont {De~Angelis}, \citenamefont {Jim{\'{e}}nez},\ and\
  \citenamefont {Casciola}}]{Cimarelli2016}%
  \BibitemOpen
  \bibfield  {author} {\bibinfo {author} {\bibfnamefont {A.}~\bibnamefont
  {Cimarelli}}, \bibinfo {author} {\bibfnamefont {E.}~\bibnamefont
  {De~Angelis}}, \bibinfo {author} {\bibfnamefont {J.}~\bibnamefont
  {Jim{\'{e}}nez}}, \ and\ \bibinfo {author} {\bibfnamefont {C.~M.}\
  \bibnamefont {Casciola}},\ }\bibfield  {title} {{\selectlanguage
  {English}\enquote {\bibinfo {title} {{Cascades and wall-normal fluxes in
  turbulent channel flows}},}\ }}\href {\doibase 10.1017/jfm.2016.275}
  {\bibfield  {journal} {\bibinfo  {journal} {Journal of Fluid Mechanics}\
  }\textbf {\bibinfo {volume} {796}},\ \bibinfo {pages} {417--436} (\bibinfo
  {year} {2016})}\BibitemShut {NoStop}%
\bibitem [{\citenamefont {Piomelli}\ \emph {et~al.}(1991)\citenamefont
  {Piomelli}, \citenamefont {Cabot}, \citenamefont {Moin},\ and\ \citenamefont
  {Lee}}]{Piomelli1991}%
  \BibitemOpen
  \bibfield  {author} {\bibinfo {author} {\bibfnamefont {U.}~\bibnamefont
  {Piomelli}}, \bibinfo {author} {\bibfnamefont {W.~H.}\ \bibnamefont {Cabot}},
  \bibinfo {author} {\bibfnamefont {P.}~\bibnamefont {Moin}}, \ and\ \bibinfo
  {author} {\bibfnamefont {S.}~\bibnamefont {Lee}},\ }\bibfield  {title}
  {\enquote {\bibinfo {title} {{Subgrid-scale backscatter in turbulent and
  transitional flows}},}\ }\href {\doibase 10.1063/1.857956} {\bibfield
  {journal} {\bibinfo  {journal} {Physics of Fluids A}\ }\textbf {\bibinfo
  {volume} {3}},\ \bibinfo {pages} {1766--1771} (\bibinfo {year}
  {1991})}\BibitemShut {NoStop}%
\bibitem [{\citenamefont {Domaradzki}, \citenamefont {Liu},\ and\ \citenamefont
  {Brachet}(1993)}]{Domaradzki1993}%
  \BibitemOpen
  \bibfield  {author} {\bibinfo {author} {\bibfnamefont {J.~A.}\ \bibnamefont
  {Domaradzki}}, \bibinfo {author} {\bibfnamefont {W.}~\bibnamefont {Liu}}, \
  and\ \bibinfo {author} {\bibfnamefont {M.~E.}\ \bibnamefont {Brachet}},\
  }\bibfield  {title} {\enquote {\bibinfo {title} {{An analysis of
  subgrid-scale interactions in numerically simulated isotropic turbulence}},}\
  }\href {\doibase 10.1063/1.858850} {\bibfield  {journal} {\bibinfo  {journal}
  {Physics of Fluids A: Fluid Dynamics}\ }\textbf {\bibinfo {volume} {5}},\
  \bibinfo {pages} {1747--1759} (\bibinfo {year} {1993})}\BibitemShut {NoStop}%
\bibitem [{\citenamefont {Domaradzki}\ \emph {et~al.}(1994)\citenamefont
  {Domaradzki}, \citenamefont {Liu}, \citenamefont {H{\"{a}}rtel},\ and\
  \citenamefont {Kleiser}}]{Domaradzki1994}%
  \BibitemOpen
  \bibfield  {author} {\bibinfo {author} {\bibfnamefont {J.~A.}\ \bibnamefont
  {Domaradzki}}, \bibinfo {author} {\bibfnamefont {W.}~\bibnamefont {Liu}},
  \bibinfo {author} {\bibfnamefont {C.}~\bibnamefont {H{\"{a}}rtel}}, \ and\
  \bibinfo {author} {\bibfnamefont {L.}~\bibnamefont {Kleiser}},\ }\bibfield
  {title} {\enquote {\bibinfo {title} {{Energy transfer in numerically
  simulated wall-bounded turbulent flows}},}\ }\href {\doibase
  10.1063/1.868272} {\bibfield  {journal} {\bibinfo  {journal} {Physics of
  Fluids}\ }\textbf {\bibinfo {volume} {6}},\ \bibinfo {pages} {1583--1599}
  (\bibinfo {year} {1994})}\BibitemShut {NoStop}%
\bibitem [{\citenamefont {H{\"{a}}rtel}\ \emph {et~al.}(1994)\citenamefont
  {H{\"{a}}rtel}, \citenamefont {Kleiser}, \citenamefont {Unger},\ and\
  \citenamefont {Friedrich}}]{Haertel1994}%
  \BibitemOpen
  \bibfield  {author} {\bibinfo {author} {\bibfnamefont {C.}~\bibnamefont
  {H{\"{a}}rtel}}, \bibinfo {author} {\bibfnamefont {L.}~\bibnamefont
  {Kleiser}}, \bibinfo {author} {\bibfnamefont {F.}~\bibnamefont {Unger}}, \
  and\ \bibinfo {author} {\bibfnamefont {R.}~\bibnamefont {Friedrich}},\
  }\bibfield  {title} {\enquote {\bibinfo {title} {{Subgrid-scale energy
  transfer in the near-wall region of turbulent flows}},}\ }\href {\doibase
  10.1063/1.868137} {\bibfield  {journal} {\bibinfo  {journal} {Physics of
  Fluids}\ }\textbf {\bibinfo {volume} {6}},\ \bibinfo {pages} {3130--3143}
  (\bibinfo {year} {1994})}\BibitemShut {NoStop}%
\bibitem [{\citenamefont {Kerr}, \citenamefont {Domaradzki},\ and\
  \citenamefont {Barbier}(1996)}]{Kerr1996}%
  \BibitemOpen
  \bibfield  {author} {\bibinfo {author} {\bibfnamefont {R.~M.}\ \bibnamefont
  {Kerr}}, \bibinfo {author} {\bibfnamefont {J.~A.}\ \bibnamefont
  {Domaradzki}}, \ and\ \bibinfo {author} {\bibfnamefont {G.}~\bibnamefont
  {Barbier}},\ }\bibfield  {title} {\enquote {\bibinfo {title} {{Small-scale
  properties of nonlinear interactions and subgrid-scale energy transfer in
  isotropic turbulence}},}\ }\href {\doibase 10.1063/1.868827} {\bibfield
  {journal} {\bibinfo  {journal} {Physics of Fluids}\ }\textbf {\bibinfo
  {volume} {8}},\ \bibinfo {pages} {197--208} (\bibinfo {year}
  {1996})}\BibitemShut {NoStop}%
\bibitem [{\citenamefont {Meneveau}(1991{\natexlab{a}})}]{Meneveau1991a}%
  \BibitemOpen
  \bibfield  {author} {\bibinfo {author} {\bibfnamefont {C.}~\bibnamefont
  {Meneveau}},\ }\bibfield  {title} {\enquote {\bibinfo {title} {{Dual spectra
  and mixed energy cascade of turbulence in the wavelet representation}},}\
  }\href {\doibase 10.1103/PhysRevLett.66.1450} {\bibfield  {journal} {\bibinfo
   {journal} {Physical Review Letters}\ }\textbf {\bibinfo {volume} {66}},\
  \bibinfo {pages} {1450--1453} (\bibinfo {year}
  {1991}{\natexlab{a}})}\BibitemShut {NoStop}%
\bibitem [{\citenamefont {Meneveau}(1991{\natexlab{b}})}]{Meneveau1991b}%
  \BibitemOpen
  \bibfield  {author} {\bibinfo {author} {\bibfnamefont {C.}~\bibnamefont
  {Meneveau}},\ }\bibfield  {title} {\enquote {\bibinfo {title} {{Analysis of
  turbulence in the orthonormal wavelet representation}},}\ }\href {\doibase
  10.1017/S0022112091003786} {\bibfield  {journal} {\bibinfo  {journal}
  {Journal of Fluid Mechanics}\ }\textbf {\bibinfo {volume} {232}},\ \bibinfo
  {pages} {469} (\bibinfo {year} {1991}{\natexlab{b}})}\BibitemShut {NoStop}%
\bibitem [{\citenamefont {L{\'{o}}pez}\ \emph {et~al.}(2020)\citenamefont
  {L{\'{o}}pez}, \citenamefont {Feldmann}, \citenamefont {Rampp}, \citenamefont
  {Vela-Mart{\'{i}}n}, \citenamefont {Shi},\ and\ \citenamefont
  {Avila}}]{Lopez2020}%
  \BibitemOpen
  \bibfield  {author} {\bibinfo {author} {\bibfnamefont {J.~M.}\ \bibnamefont
  {L{\'{o}}pez}}, \bibinfo {author} {\bibfnamefont {D.}~\bibnamefont
  {Feldmann}}, \bibinfo {author} {\bibfnamefont {M.}~\bibnamefont {Rampp}},
  \bibinfo {author} {\bibfnamefont {A.}~\bibnamefont {Vela-Mart{\'{i}}n}},
  \bibinfo {author} {\bibfnamefont {L.}~\bibnamefont {Shi}}, \ and\ \bibinfo
  {author} {\bibfnamefont {M.}~\bibnamefont {Avila}},\ }\bibfield  {title}
  {\enquote {\bibinfo {title} {{nsCouette – A high-performance code for
  direct numerical simulations of turbulent Taylor–Couette flow}},}\ }\href
  {\doibase 10.1016/j.softx.2019.100395} {\bibfield  {journal} {\bibinfo
  {journal} {SoftwareX}\ }\textbf {\bibinfo {volume} {11}},\ \bibinfo {pages}
  {100395} (\bibinfo {year} {2020})}\BibitemShut {NoStop}%
\bibitem [{\citenamefont {Wu}, \citenamefont {Baltzer},\ and\ \citenamefont
  {Adrian}(2012)}]{Wu2012}%
  \BibitemOpen
  \bibfield  {author} {\bibinfo {author} {\bibfnamefont {X.}~\bibnamefont
  {Wu}}, \bibinfo {author} {\bibfnamefont {J.~R.}\ \bibnamefont {Baltzer}}, \
  and\ \bibinfo {author} {\bibfnamefont {R.~J.}\ \bibnamefont {Adrian}},\
  }\bibfield  {title} {\enquote {\bibinfo {title} {{Direct numerical simulation
  of a 30R long turbulent pipe flow at R + = 685: Large-and very large-scale
  motions}},}\ }\href {\doibase 10.1017/jfm.2012.81} {\bibfield  {journal}
  {\bibinfo  {journal} {Journal of Fluid Mechanics}\ }\textbf {\bibinfo
  {volume} {698}},\ \bibinfo {pages} {235--281} (\bibinfo {year}
  {2012})}\BibitemShut {NoStop}%
\bibitem [{\citenamefont {Kelley}, \citenamefont {Allshouse},\ and\
  \citenamefont {Ouellette}(2013)}]{Kelley2013}%
  \BibitemOpen
  \bibfield  {author} {\bibinfo {author} {\bibfnamefont {D.~H.}\ \bibnamefont
  {Kelley}}, \bibinfo {author} {\bibfnamefont {M.~R.}\ \bibnamefont
  {Allshouse}}, \ and\ \bibinfo {author} {\bibfnamefont {N.~T.}\ \bibnamefont
  {Ouellette}},\ }\bibfield  {title} {{\selectlanguage {English}\enquote
  {\bibinfo {title} {{Lagrangian coherent structures separate dynamically
  distinct regions in fluid flows}},}\ }}\href {\doibase
  10.1103/PhysRevE.88.013017} {\bibfield  {journal} {\bibinfo  {journal}
  {Physical Review E - Statistical, Nonlinear, and Soft Matter Physics}\
  }\textbf {\bibinfo {volume} {88}},\ \bibinfo {pages} {3--6} (\bibinfo {year}
  {2013})}\BibitemShut {NoStop}%
\bibitem [{\citenamefont {Ballouz}\ and\ \citenamefont
  {Ouellette}(2018)}]{Ballouz2018}%
  \BibitemOpen
  \bibfield  {author} {\bibinfo {author} {\bibfnamefont {J.~G.}\ \bibnamefont
  {Ballouz}}\ and\ \bibinfo {author} {\bibfnamefont {N.~T.}\ \bibnamefont
  {Ouellette}},\ }\bibfield  {title} {\enquote {\bibinfo {title} {{Tensor
  geometry in the turbulent cascade}},}\ }\href {\doibase 10.1017/jfm.2017.802}
  {\bibfield  {journal} {\bibinfo  {journal} {Journal of Fluid Mechanics}\
  }\textbf {\bibinfo {volume} {835}},\ \bibinfo {pages} {1048--1064} (\bibinfo
  {year} {2018})}\BibitemShut {NoStop}%
\bibitem [{\citenamefont {Feldmann}, \citenamefont {Bauer},\ and\ \citenamefont
  {Wagner}(2018)}]{Feldmann2018}%
  \BibitemOpen
  \bibfield  {author} {\bibinfo {author} {\bibfnamefont {D.}~\bibnamefont
  {Feldmann}}, \bibinfo {author} {\bibfnamefont {C.}~\bibnamefont {Bauer}}, \
  and\ \bibinfo {author} {\bibfnamefont {C.}~\bibnamefont {Wagner}},\
  }\bibfield  {title} {\enquote {\bibinfo {title} {{Computational domain length
  and Reynolds number effects on large-scale coherent motions in turbulent pipe
  flow}},}\ }\href {\doibase 10.1080/14685248.2017.1418086} {\bibfield
  {journal} {\bibinfo  {journal} {Journal of Turbulence}\ }\textbf {\bibinfo
  {volume} {19}},\ \bibinfo {pages} {274--295} (\bibinfo {year}
  {2018})}\BibitemShut {NoStop}%
\bibitem [{\citenamefont {Bauer}, \citenamefont {Feldmann},\ and\ \citenamefont
  {Wagner}(2017)}]{Bauer2017}%
  \BibitemOpen
  \bibfield  {author} {\bibinfo {author} {\bibfnamefont {C.}~\bibnamefont
  {Bauer}}, \bibinfo {author} {\bibfnamefont {D.}~\bibnamefont {Feldmann}}, \
  and\ \bibinfo {author} {\bibfnamefont {C.}~\bibnamefont {Wagner}},\
  }\bibfield  {title} {\enquote {\bibinfo {title} {{On the convergence and
  scaling of high-order statistical moments in turbulent pipe flow using direct
  numerical simulations}},}\ }\href {\doibase 10.1063/1.4996882} {\bibfield
  {journal} {\bibinfo  {journal} {Physics of Fluids}\ }\textbf {\bibinfo
  {volume} {29}} (\bibinfo {year} {2017}),\ 10.1063/1.4996882}\BibitemShut
  {NoStop}%
\bibitem [{\citenamefont {Buzzicotti}\ \emph
  {et~al.}(2018{\natexlab{b}})\citenamefont {Buzzicotti}, \citenamefont
  {Linkmann}, \citenamefont {Aluie}, \citenamefont {Biferale}, \citenamefont
  {Brasseur},\ and\ \citenamefont {Meneveau}}]{Buzzicotti2018}%
  \BibitemOpen
  \bibfield  {author} {\bibinfo {author} {\bibfnamefont {M.}~\bibnamefont
  {Buzzicotti}}, \bibinfo {author} {\bibfnamefont {M.}~\bibnamefont
  {Linkmann}}, \bibinfo {author} {\bibfnamefont {H.}~\bibnamefont {Aluie}},
  \bibinfo {author} {\bibfnamefont {L.}~\bibnamefont {Biferale}}, \bibinfo
  {author} {\bibfnamefont {J.}~\bibnamefont {Brasseur}}, \ and\ \bibinfo
  {author} {\bibfnamefont {C.}~\bibnamefont {Meneveau}},\ }\bibfield  {title}
  {\enquote {\bibinfo {title} {{Effect of filter type on the statistics of
  energy transfer between resolved and subfilter scales from a-priori analysis
  of direct numerical simulations of isotropic turbulence}},}\ }\href {\doibase
  10.1080/14685248.2017.1417597} {\bibfield  {journal} {\bibinfo  {journal}
  {Journal of Turbulence}\ }\textbf {\bibinfo {volume} {19}},\ \bibinfo {pages}
  {167--197} (\bibinfo {year} {2018}{\natexlab{b}})}\BibitemShut {NoStop}%
\bibitem [{\citenamefont {Leslie}\ and\ \citenamefont
  {Quarini}(1979)}]{Leslie1979}%
  \BibitemOpen
  \bibfield  {author} {\bibinfo {author} {\bibfnamefont {D.~C.}\ \bibnamefont
  {Leslie}}\ and\ \bibinfo {author} {\bibfnamefont {G.~L.}\ \bibnamefont
  {Quarini}},\ }\bibfield  {title} {\enquote {\bibinfo {title} {{The
  application of turbulence theory to the formulation of subgrid modelling
  procedures}},}\ }\href {\doibase 10.1017/S0022112079000045} {\bibfield
  {journal} {\bibinfo  {journal} {Journal of Fluid Mechanics}\ }\textbf
  {\bibinfo {volume} {91}},\ \bibinfo {pages} {65} (\bibinfo {year}
  {1979})}\BibitemShut {NoStop}%
\bibitem [{\citenamefont {{Jan Chen}}(2019)}]{Chen2019}%
  \BibitemOpen
  \bibfield  {author} {\bibinfo {author} {\bibnamefont {{Jan Chen}}},\ }\emph
  {\bibinfo {title} {{Interscale Energy Flux in Wall-Bounded Turbulent
  Flow}}},\ \href@noop {} {Ph.D. thesis},\ \bibinfo  {school} {Carl von
  Ossietzky Universit{\"{a}}t}, \bibinfo {address} {Oldenburg} (\bibinfo {year}
  {2019})\BibitemShut {NoStop}%
\bibitem [{\citenamefont {Aoyama}\ \emph {et~al.}(2005)\citenamefont {Aoyama},
  \citenamefont {Ishihara}, \citenamefont {Kaneda}, \citenamefont {Yokokawa},
  \citenamefont {Itakura},\ and\ \citenamefont {Uno}}]{Aoyama2005}%
  \BibitemOpen
  \bibfield  {author} {\bibinfo {author} {\bibfnamefont {T.}~\bibnamefont
  {Aoyama}}, \bibinfo {author} {\bibfnamefont {T.}~\bibnamefont {Ishihara}},
  \bibinfo {author} {\bibfnamefont {Y.}~\bibnamefont {Kaneda}}, \bibinfo
  {author} {\bibfnamefont {M.}~\bibnamefont {Yokokawa}}, \bibinfo {author}
  {\bibfnamefont {K.}~\bibnamefont {Itakura}}, \ and\ \bibinfo {author}
  {\bibfnamefont {A.}~\bibnamefont {Uno}},\ }\bibfield  {title} {\enquote
  {\bibinfo {title} {{Statistics of energy transfer in high-resolution direct
  numerical simulation of turbulence in a periodic box}},}\ }\href {\doibase
  10.1143/JPSJ.74.3202} {\bibfield  {journal} {\bibinfo  {journal} {Journal of
  the Physical Society of Japan}\ }\textbf {\bibinfo {volume} {74}},\ \bibinfo
  {pages} {3202--3212} (\bibinfo {year} {2005})}\BibitemShut {NoStop}%
\bibitem [{\citenamefont {Gonzalez}\ and\ \citenamefont
  {Woods}(2002)}]{Gonzalez2002}%
  \BibitemOpen
  \bibfield  {author} {\bibinfo {author} {\bibfnamefont {R.~C.}\ \bibnamefont
  {Gonzalez}}\ and\ \bibinfo {author} {\bibfnamefont {R.~E.}\ \bibnamefont
  {Woods}},\ }\href@noop {} {\emph {\bibinfo {title} {{Digital Image
  Processing}}}}\ (\bibinfo  {publisher} {Prentice-Hall},\ \bibinfo {year}
  {2002})\BibitemShut {NoStop}%
\bibitem [{\citenamefont {Von~Kameke}\ \emph {et~al.}(2011)\citenamefont
  {Von~Kameke}, \citenamefont {Huhn}, \citenamefont
  {Fern{\'{a}}ndez-Garc{\'{i}}a}, \citenamefont {Mu{\~{n}}uzuri},\ and\
  \citenamefont {P{\'{e}}rez-Mu{\~{n}}uzuri}}]{vonKameke2011}%
  \BibitemOpen
  \bibfield  {author} {\bibinfo {author} {\bibfnamefont {A.}~\bibnamefont
  {Von~Kameke}}, \bibinfo {author} {\bibfnamefont {F.}~\bibnamefont {Huhn}},
  \bibinfo {author} {\bibfnamefont {G.}~\bibnamefont
  {Fern{\'{a}}ndez-Garc{\'{i}}a}}, \bibinfo {author} {\bibfnamefont {A.~P.}\
  \bibnamefont {Mu{\~{n}}uzuri}}, \ and\ \bibinfo {author} {\bibfnamefont
  {V.}~\bibnamefont {P{\'{e}}rez-Mu{\~{n}}uzuri}},\ }\bibfield  {title}
  {\enquote {\bibinfo {title} {{Double Cascade Turbulence and Richardson
  Dispersion in a Horizontal Fluid Flow Induced by Faraday Waves}},}\ }\href
  {\doibase 10.1103/PhysRevLett.107.074502} {\bibfield  {journal} {\bibinfo
  {journal} {Physical Review Letters}\ }\textbf {\bibinfo {volume} {107}},\
  \bibinfo {pages} {1--4} (\bibinfo {year} {2011})}\BibitemShut {NoStop}%
\bibitem [{\citenamefont {Xia}\ \emph {et~al.}(2011)\citenamefont {Xia},
  \citenamefont {Byrne}, \citenamefont {Falkovich},\ and\ \citenamefont
  {Shats}}]{Xia2011}%
  \BibitemOpen
  \bibfield  {author} {\bibinfo {author} {\bibfnamefont {H.}~\bibnamefont
  {Xia}}, \bibinfo {author} {\bibfnamefont {D.}~\bibnamefont {Byrne}}, \bibinfo
  {author} {\bibfnamefont {G.}~\bibnamefont {Falkovich}}, \ and\ \bibinfo
  {author} {\bibfnamefont {M.}~\bibnamefont {Shats}},\ }\bibfield  {title}
  {\enquote {\bibinfo {title} {{Upscale energy transfer in thick turbulent
  fluid layers}},}\ }\href {\doibase 10.1038/nphys1910} {\bibfield  {journal}
  {\bibinfo  {journal} {Nature Physics}\ }\textbf {\bibinfo {volume} {7}},\
  \bibinfo {pages} {321--324} (\bibinfo {year} {2011})}\BibitemShut {NoStop}%
\bibitem [{\citenamefont {{\"{O}}rl{\"{u}}}\ and\ \citenamefont
  {Schlatter}(2011)}]{Oerlue2011}%
  \BibitemOpen
  \bibfield  {author} {\bibinfo {author} {\bibfnamefont {R.}~\bibnamefont
  {{\"{O}}rl{\"{u}}}}\ and\ \bibinfo {author} {\bibfnamefont {P.}~\bibnamefont
  {Schlatter}},\ }\bibfield  {title} {\enquote {\bibinfo {title} {{On the
  fluctuating wall-shear stress in zero pressure-gradient turbulent boundary
  layer flows}},}\ }\href {\doibase 10.1063/1.3555191} {\bibfield  {journal}
  {\bibinfo  {journal} {Physics of Fluids}\ }\textbf {\bibinfo {volume} {23}}
  (\bibinfo {year} {2011}),\ 10.1063/1.3555191}\BibitemShut {NoStop}%
\bibitem [{\citenamefont {Vreman}, \citenamefont {Geurts},\ and\ \citenamefont
  {Kuerten}(1994)}]{Vreman1994}%
  \BibitemOpen
  \bibfield  {author} {\bibinfo {author} {\bibfnamefont {B.}~\bibnamefont
  {Vreman}}, \bibinfo {author} {\bibfnamefont {B.}~\bibnamefont {Geurts}}, \
  and\ \bibinfo {author} {\bibfnamefont {H.}~\bibnamefont {Kuerten}},\
  }\bibfield  {title} {\enquote {\bibinfo {title} {{Realizability Conditions
  for the Turbulent Stress Tensor in Large-Eddy Simulation}},}\ }\href
  {\doibase 10.1017/S0022112094003745} {\bibfield  {journal} {\bibinfo
  {journal} {Journal of Fluid Mechanics}\ }\textbf {\bibinfo {volume} {278}},\
  \bibinfo {pages} {351--362} (\bibinfo {year} {1994})}\BibitemShut {NoStop}%
\bibitem [{\citenamefont {Sadek}\ and\ \citenamefont
  {Aluie}(2018)}]{Sadek2018}%
  \BibitemOpen
  \bibfield  {author} {\bibinfo {author} {\bibfnamefont {M.}~\bibnamefont
  {Sadek}}\ and\ \bibinfo {author} {\bibfnamefont {H.}~\bibnamefont {Aluie}},\
  }\bibfield  {title} {\enquote {\bibinfo {title} {{Extracting the spectrum of
  a flow by spatial filtering}},}\ }\href {\doibase
  10.1103/PhysRevFluids.3.124610} {\bibfield  {journal} {\bibinfo  {journal}
  {Physical Review Fluids}\ }\textbf {\bibinfo {volume} {3}},\ \bibinfo {pages}
  {124610} (\bibinfo {year} {2018})}\BibitemShut {NoStop}%
\bibitem [{\citenamefont {Pope}(2000)}]{Pope2000}%
  \BibitemOpen
  \bibfield  {author} {\bibinfo {author} {\bibfnamefont {S.~B.}\ \bibnamefont
  {Pope}},\ }\href@noop {} {\emph {\bibinfo {title} {{Turbulent Flows}}}},\
  \bibinfo {edition} {7th}\ ed.\ (\bibinfo  {publisher} {Cambridge University
  Press},\ \bibinfo {year} {2000})\ p.\ \bibinfo {pages} {771}\BibitemShut
  {NoStop}%
\end{thebibliography}%

\appendix

\section{Filter kernel}
\label{app:filterKernel}

Following \textcite{Pope2000}, we define the three one-dimensional filter kernels
\begin{align}
\text{sharp spectral:}\qquad
G^{\lambda}(x) &= \operatorname{sinc}\!\left(x\frac{\pi}{\lambda}\right)\text{,}\label{eq:kernelFourierPhys}\\
\hat{G}^{\lambda}(\kappa) &= \operatorname{H}\!\left(\frac{\pi}{\lambda}-\lvert\kappa\rvert\right)\text{,}\label{eq:kernelFourier}\\
\text{Gaussian:}\qquad
G^{\lambda}(x) &= \sqrt{\frac{6}{\pi\lambda^2}}\exp\!\left(\frac{-6 x^2}{\lambda^2}\right)\text{,}\label{eq:kernelGaussPhys}\\
\hat{G}^{\lambda}(\kappa) &= \exp\!\left(\frac{-\lambda^2\kappa^2}{24}\right)\text{,}\label{eq:kernelGauss}\\
\text{box}\qquad
G^{\lambda}(x) &= \frac{1}{\lambda}\operatorname{H}\!\left(\frac{\lambda}{2}-\lvert x\rvert\right)\text{,}\label{eq:kernelBoxPhys}\\
\hat{G}^{\lambda}(\kappa) &= \operatorname{sinc}\!\left(\kappa\frac{\lambda}{2}\right)\label{eq:kernelBox}\text{,}
\end{align}
where $G^{\lambda}$ denotes the filter kernel in physical space ($x$) and $\hat{G}^{\lambda}$ the filter kernel in wavenumber space ($\kappa$); $\lambda$ denotes the nominal filter scale. Moreover, $\operatorname{sinc}$ refers to the unnormalised cardinal sine function, and $\operatorname{H}$ to the Heaviside step function.

Note that any of the three kernels can be applied either in physical space---via convolution with $G$---or in spectral space multiplication with $\hat{G}$. Both routes are mathematically equivalent and preserve the intrinsic properties of the chosen kernel. The sharp spectral kernel exhibits compact spectral support by strictly truncating high wavenumbers, which renders it strongly non-local in physical space due to the slow (algebraic) decay of its kernel. In contrast, the box kernel is compact in physical space but possesses slowly decaying spectral characteristics, leading to poor scale separation in wavenumber space. The Gaussian kernel, with its exponential decay in both domains, provides the best compromise in terms of localisation, albeit without compact support in either space.

The actual 2D filter kernel we apply to eq.~\eqref{eq:explicitFilterFourierSpace}, is constructed by the outer product of two 1D kernels as
\begin{align}
\widehat{G}^{\lambda_{\theta}\times\lambda_{z}}\left(\kappa_{\theta}, \kappa_{z}\right) = 
\widehat{G}^{\lambda_{\theta}}\left(\kappa_{\theta}\right) \otimes
\widehat{G}^{\lambda_{z}}\left(\kappa_{z}\right)
\text{.}
\label{eq:2dFilterKernel}
\end{align}

\section{Statistical analysis}
\label{app:statisticalAnalysis}

To quantify the effect of the filter kernel on the local structure of the inter-scale energy flux statistically, we compute typical one-point auto-correlations
\begin{align}
\text{RMS:}&\quad
\lla\alpha^{\prime}\alpha^{\prime}\rra^{\sfrac{1}{2}}
\text{,}\label{eq:rms}\\
\text{Skewness:}&\quad
\frac{\lla\alpha^{\prime}\alpha^{\prime}\alpha^{\prime}\rra}
{\lla\alpha^{\prime}\alpha^{\prime}\rra^{\sfrac{3}{2}}}
\text{,}\label{eq:skewness}\\
\text{Flatness:}&\quad
\frac{\lla\alpha^{\prime}\alpha^{\prime}\alpha^{\prime}\alpha^{\prime}\rra}
{\lla\alpha^{\prime}\alpha^{\prime}\rra^{\sfrac{4}{2}}}
\label{eq:flatness}
\end{align}
for $\alpha=\Pi$ at all wall-normal locations, as presented in section~\ref{sec:onePointStatisticsFourier} and section~\ref{sec:onePointStatisticsCompare}. The angled brackets and the prime superscript denote mean and fluctuating quantities analogously to eq.~\eqref{eq:reynoldsDecomposition}.

Additionally, we analyse the 1D and 2D two-point correlation functions
\begin{align}
C_{\alpha\beta}\left(r_{0}, \Delta\theta\right)& =
\frac{\lla \alpha\left(r_{0}, \theta_{0}, z_{0}, t \right)
      \cdot \beta\left(r_{0}, \theta_{0}+\Delta\theta, z_{0}, t\right)\rra}
{\lla\alpha^{\prime}\beta^{\prime}\rra}
\label{eq:twoPointStatisticsTh}\\
C_{\alpha\beta}\left(r_{0}, \Delta z \right)& =
\frac{\lla \alpha\left(r_{0}, \theta_{0}, z_{0}, t \right)
      \cdot \beta\left(r_{0}, \theta_{0}, z_{0}+\Delta z, t\right)\rra}
{\lla\alpha^{\prime}\beta^{\prime}\rra}
\label{eq:twoPointStatisticsZ}\\
C_{\alpha\beta}\left(r_{0}, \Delta\theta, \Delta z \right)& =
\frac{\lla \alpha\left(r_{0}, \theta_{0}, z_{0}, t \right)
      \cdot \beta\left(r_{0}, \theta_{0}+\Delta\theta, z_{0}+\Delta z, t\right)\rra}
{\lla\alpha^{\prime}\beta^{\prime}\rra}
\label{eq:twoPointStatisticsThZ}
\end{align}
at a fixed radial location $r_{0}$. In this study, we focus on one single location in the buffer layer ($r_0=\num{0.93}R$, $r_0^+=168$), that corresponds to a wall-normal distance of $y^+=(R-r)\,\ReTau=12$. In order to better interpret azimuthal correlations in terms of physical lengths (\eg the width or spacing of streaks), we convert the angular separation, $\Delta\theta$, into a physical length scale, $\Delta\theta r^+$, for all correlation plots shown here.

The quantities $\alpha$ and $\beta$ can be the energy flux $\Pi^\lambda$, according to eq.~\eqref{eq:eFlux}, the streamwise velocity fluctuations $u_z'$, according to eq.~\eqref{eq:reynoldsDecomposition}, the streamwise vorticity
\begin{align}
\omega_{z} = \frac{1}{r}\left(\frac{\partial\left(r u_{\theta}\right)}{\partial r} -
\frac{\partial u_{r}}{\partial\theta}\right)\text{,}
\label{eq:vorticity}
\end{align}
or one of the field variables representing typical events according to the quadrant analysis detailed in Appendix~\ref{app:quadrantAnalysis}.

\section{Quadrant analysis\label{app:quadrantAnalysis}}

Besides streamwise streaks ($u^{\prime}_{z}$) and streamwise-aligned vortices ($\omega_{z}$), another class of structural features, that play an important role in the turbulence near-wall cycle, are so-called $Q$ events \cite{Wallace2016}. A sweep event ($Q_{4}$) represents movement of high-speed fluid towards the wall, while an ejection event ($Q_{2}$) is detected where low-speed fluid is moving away from the wall. Therefore, sweeps are most often associated with high-speed streaks and ejections with low-speed streaks. An inward interaction ($Q_{3}$) represents movement of low-speed fluid towards the wall, while an outward interaction ($Q_{1}$) is detected where high-speed fluid is moving away from the wall.

In order to compute cross-correlations (see eqs.~\eqref{eq:twoPointStatisticsTh} to \eqref{eq:twoPointStatisticsThZ}) between the energy flux field and localised $Q$ events, we extract the instantaneous quantities
\begin{align}
Q_{1}&=\begin{cases}u_{r}u_{z}\text{~if~}u_{r}<0\land u_{z}>0\\0\text{~otherwise}\end{cases}\\
Q_{2}&=\begin{cases}u_{r}u_{z}\text{~if~}u_{r}<0\land u_{z}<0\\0\text{~otherwise}\end{cases}\\
Q_{3}&=\begin{cases}u_{r}u_{z}\text{~if~}u_{r}>0\land u_{z}<0\\0\text{~otherwise}\end{cases}\\
Q_{4}&=\begin{cases}u_{r}u_{z}\text{~if~}u_{r}>0\land u_{z}>0\\0\text{~otherwise}\end{cases}
\label{eq:qEvents}
\end{align}
from the turbulent velocity fields.

Note, that---in contrast to the usual convention used for $Q$ events---here a positive wall-normal velocity component ($u_r$) denotes movement towards the wall, and not away from the wall, since we use a cylindrical co-ordinate system ($r$, $\theta$, $z$) throughout the paper to describe the pipe flow. Therefore, the scalar fields $Q_2$ and $Q_4$ are purely positive, whereas $Q_1$ and $Q_3$ are purely negative.

\end{document}